# Hardening Soft Information: Evidence on Analyst Integration Costs

**Kerry Xiao**

Hong Kong University of Science and Technology

ackerry@ust.hk

**Amy Zang***

Hong Kong University of Science and Technology

amy.zang@ust.hk

September 2026

## Abstract

We examine how the cost of transforming qualitative information into precise numerical estimates—a form of integration cost—creates a structural friction in expectations formation. To isolate this integration cost from the costs of information awareness and acquisition, we exploit sell-side analyst reports, in which the same forecaster simultaneously produces textual narratives and numerical forecasts. Because the information underlying the text has already been acquired, any systematic gap between the two outputs can be attributed to integration costs. We document systematic quantification inefficiency: an analyst's textual tone negatively predicts her contemporaneous forecast errors and positively predicts her subsequent numerical revisions, revealing that analysts leave part of their qualitative insights unquantified until further evidence arrives. Consistent with this integration-friction explanation, this inefficiency intensifies when reports are linguistically vaguer, environmental uncertainty is higher, or analysts' processing capacity is more constrained, and it persists where strategic and behavioral explanations are weaker. Our findings provide direct, large-sample evidence that integration costs constitute a distinct economic friction, explaining why soft information carries value-relevant content beyond contemporaneous hard numbers.

**Keywords:** information frictions, information processing costs, integration costs, expectations formation, analyst forecasts, soft information, textual analysis

**JEL classification:** D83; D84; G14; G24; M41

* Corresponding author.

This paper evolved from earlier work titled "Hardening Soft Information: Analyst Conservative Bias" and "Vague Knowledge: Evidence from Analyst Reports." We are grateful for helpful comments from Andrew Van Buskirk (discussant), Allen Huang, Irene Karamanou (discussant), Roby Lehavy, Tim Loughran, Jedson Pinto (discussant), Guoman She, Haifeng You, Frank Yu, Tianyu Zhang, and Rong Zheng; from seminar participants at HKUST, HKU, Tsinghua, and CEIBS; and from conference participants at CUHK Textual Analysis Conference 2017, Utah Winter Accounting Conference 2018, EAA 2018, AAA 2024, and the Mediterranean Accounting Conference 2025. All errors are our own.

## Hardening Soft Information: Evidence on Analyst Integration Costs

## 1. Introduction

Efficient information processing by capital market participants underpins price discovery and capital allocation. A cornerstone of modern disclosure theory is that information, even when publicly disclosed, is not frictionlessly incorporated into market prices. In their seminal review, Blankespoor et al. (2020) synthesize the extensive literature into a multi-stage taxonomy of information processing costs, identifying: (i) the cost of becoming aware that a specific disclosure exists, (ii) the cost of acquiring the raw information encoded within that disclosure, and (iii) the cost of integrating those acquired signals into a structured format for valuation decisions. While this structural framework has profoundly shaped our theoretical understanding of information frictions, establishing clean empirical identification remains a primary challenge.

In most archival settings, an information user's awareness, acquisition, and integration costs tightly co-vary, obscuring the independent economic consequences of any single processing friction. This study addresses this identification challenge by exploiting a unique, naturally occurring laboratory: the qualitative and quantitative outputs that are simultaneously produced in research reports by sell-side equity analysts. We provide the first large-sample evidence that isolates integration costs—specifically, the friction of transforming an agent's own acquired qualitative, narrative knowledge into precise quantitative projections—and demonstrate how these costs lead to systematic "quantification inefficiency" in the expectations-formation process of sophisticated market participants.

Sell-side analyst reports provide an ideal empirical setting to isolate integration costs because they contain two distinct, simultaneously produced output formats: numerical forecasts (e.g., earnings forecasts, target prices, stock recommendations) and textual narratives. The numerical output captures the analyst's final, integrated judgment expressed as a precise estimate. In contrast, the accompanying text contains rich, multi-dimensional, and frequently vague descriptions of firm fundamentals, such as firm-specific strategic positioning, changes in the competitive landscape, and qualitative assessments of management quality. Crucially, this institutional setting creates a clean separation between raw signal acquisition and final information integration. Because the information embedded within an analyst's written text has already been acquired, the acquisition costs at the stage of integrating that information into her own numerical forecasts are zero by construction. However, the cognitive and economic cost of integrating that unstructured qualitative knowledge into an exact numerical forecast can be substantial. This processing barrier, which we conceptualize as a quantification cost, is one type of integration cost. It stems from the cognitive effort required to interpret vague signals, assign weights to highly subjective evidence, and resolve Knightian uncertainty involved in translating narrative content into structured corporate financial metrics.[1]

When translating text into numbers is costly, a rational, capacity-constrained analyst faces a trade-off. She weighs the marginal benefit of expending scarce cognitive resources to fully calculate and embed every value-relevant implication of her qualitative text into her current forecast against the opportunity cost of her limited time and attention. When integration costs are high, the analyst rationally chooses not to quantify everything and leaves a portion of her acquired signals unquantified, communicating those insights exclusively through the textual narrative. This rational inattention results in *quantification inefficiency*, whereby the analyst's own numerical forecasts systematically underweight the directional implications of the qualitative insights she has already conveyed textually.

[1] Knightian uncertainty refers to situations where underlying probabilities cannot be quantified. Knight (1921) distinguishes *risk*, where the possible outcomes are known and their probabilities are estimable, from *uncertainty*, where no probability distribution exists over future states and expected-value calculations are infeasible.

This rational inattention model yields our incomplete quantification hypothesis (H1), which predicts that due to quantification costs, an analyst's contemporaneous earnings forecast will fail to fully incorporate the directional implications of her own report text. Specifically, when an analyst writes an optimistic report—conveying positive qualitative fundamental news she has acquired—her contemporaneous numerical forecast tends to be lower than realized earnings; conversely, a pessimistic report text, which reflects negative qualitative news, is systematically linked to an accompanying forecast that is too high relative to realized earnings. Empirically, this quantification friction predicts a negative association between machine-extracted textual tone and the contemporaneous forecast error (measured as forecast minus actual earnings).

Beyond these static consequences, our model yields a dynamic prediction regarding the expectations-updating process, which we term the delayed quantification hypothesis (H2). The unquantified value-relevant information embedded within the analyst's report text represents a latent fundamental signal. As time elapses, the analyst receives additional, more precise public information (e.g., subsequent corporate disclosures, macroeconomic indicators, or peer competitor actions) or dedicates further research effort to the covered firm. This progressive accumulation of data increases the signal-to-noise ratio of the previously unquantified information, leading to a predictable, gradual quantification path. Consequently, we hypothesize that the tone of an analyst's report text will positively predict her subsequent revisions to earnings forecasts, target prices, and stock recommendations for the same firm. Information initially confined to the text because of high quantification costs flows into revised numerical forecasts as quantification friction gradually eases.

To empirically test these hypotheses, we assemble a large sample of 241,045 analyst reports covering S&P 500 firms from 1995 to 2012 by matching Investext analyst reports to corresponding individual forecasts in the I/B/E/S database. We measure the overall directional sentiment of the narrative text (i.e., tone) using the FinBERT large language model classifier (Huang et al., 2023). FinBERT is specifically trained on a massive corpus of financial documents, enabling it to extract contextual meaning and linguistic nuance more effectively than traditional dictionary-based or generic machine-learning methods. This approach intentionally captures the directional implication of qualitative statements while remaining agnostic about their cardinal magnitude, consistent with the inherently vague nature of the narratives we study.

A key strength of our research design is that we evaluate the relation between text and numbers produced simultaneously within the same report by the same human forecaster. By construction, our research design isolates the effect of integration costs by holding both information awareness and information acquisition costs perfectly constant. Any documented systematic relation between textual tone and either *ex post* forecast errors or subsequent revisions speaks purely to the structural inefficiency at the information integration stage.

Our first set of empirical results provides strong evidence for the incomplete quantification hypothesis (H1). We document a robust, statistically significant negative relation between report-level textual tone and contemporaneous earnings forecast errors across a battery of specifications, including those with analyst-year and analyst-firm fixed effects. An optimistic textual tone predicts that the analyst's contemporaneous numerical forecast is too low relative to the subsequent earnings realization, whereas a pessimistic tone signals that the forecast is too high. In terms of economic magnitude, a one-standard-deviation increase in textual tone is associated with a decrease in the earnings forecast error of approximately 6% of its standard deviation. While this systematic forecast inefficiency is inconsistent with full-information rational expectations, it is a direct outcome of rational inattention, in which costly integration prevents human experts from fully quantifying qualitative signals.

Turning to our dynamic prediction (H2), our second set of findings confirms the presence of delayed quantification. Conditional on the contemporaneous forecast revision and standard controls, an analyst's report tone is a strong leading indicator for her future numerical revisions. More positive (negative) textual tone predicts larger upward (downward) revisions in her future earnings forecasts, target prices, and stock recommendations. Consistent with a gradual quantification process, this predictive power persists across multiple consecutive reports within the same fiscal year. These patterns align with our hypothesis: as underlying uncertainty resolves through further analyst research, additional firm disclosures, or the arrival of new public signals, analysts gradually quantify the previously unquantified narrative content into numerical forecasts.

To confirm that integration costs drive this quantification inefficiency, we implement a series of cross-sectional tests using report-level linguistic attributes and operational constraints as proxies for the degree of quantification cost.

First, we find that when a report uses vaguer language—for example, containing more sentences without numbers or more hedging words such as "maybe" or "roughly"—the negative association between report tone and forecast errors becomes two to nearly four times as strong. Because deriving numerical implications from vaguer language raises the cost of integrating, rather than acquiring, information, this pattern supports the inference that higher quantification costs drive the effect.

Second, we show that the quantification delay is strongly influenced by environmental uncertainty: the predictive power of tone for future numerical revisions is amplified under longer forecast horizons, elevated macroeconomic uncertainty (as measured by the VIX index or during recessions), higher firm-specific volatility, major disruptive corporate events (such as M&A, segment changes, or CEO/CFO turnover), and poorer firm information environments, all of which arguably increase quantification costs.

Third, consistent with capacity constraints increasing the shadow price of cognitive effort, we find that the integration friction is significantly exacerbated when analysts are busier, as proxied by larger coverage portfolios or by concurrent earnings announcements across the firms they cover.

To further support our interpretation of quantification inefficiency as an information friction, we conduct a series of tests designed to delineate our mechanism from competing strategic or behavioral explanations. First, we find that both positive and negative textual signals predict forecast errors and subsequent revisions symmetrically in their expected directions, which is inconsistent with intentional optimism bias—the concern that analysts hide negative information in text to preserve management access. Second, we repeat our main regressions in subsamples of non-all-star analysts, independent research firms, and firms with low institutional ownership, where analysts' incentives to cater to institutional clients or safeguard reputational capital are muted. The friction remains highly significant in all three groups, indicating that quantification inefficiency is not merely driven by strategic tipping of qualitative insights to preferred institutional clients or by deliberate downplaying of uncertain signals. Third, our results remain strong among highly experienced analysts. Because naive cognitive errors tend to diminish with experience and market selection, this finding indicates that the documented inefficiency represents a structural friction inherent to the integration of qualitative information.

Our paper makes important contributions to multiple strands of literature across accounting, finance, and economics. First, we advance the literature on information processing costs by providing the first large-sample empirical evidence that directly isolates integration costs. By using a within-report design that holds acquisition costs at zero by construction, we avoid the correlated cross-sectional proxies that confound acquisition with integration, demonstrating that the successful acquisition of information does not guarantee its efficient integration into asset valuations.

Second, we extend the literature on expectations formation. While prior studies using the Coibion and Gorodnichenko (2015) framework attribute individual forecasters' forecast errors predicted by *quantified* signals to nonrational expectations, we expand the information set to include observable qualitative signals. We demonstrate that analysts' forecasts systematically underreact to their own qualitative expectations, and that this underreaction is directly linked to quantification costs. Thus, we provide micro-level evidence that deviations from full-information rational expectations can arise from a previously overlooked source: the friction of integrating qualitative information.

Third, we contribute to the literature on soft information by shifting the paradigm from the information user to the *information producer*. We show that text retains incremental value beyond contemporaneous numbers not merely because it provides interpretive context, but because high quantification costs lead producers to leave some value-relevant *content* unquantified in textual form.

Finally, our study speaks to the emerging "man plus machine" literature (e.g., Cao et al., 2024) by pinpointing which human information-processing friction technology can relieve. While human analysts retain a durable advantage in discovering primary qualitative signals in the physical world and articulating them in narrative form (Liu, 2022; Christ et al., 2025), capacity constraints limit their efficiency in quantifying these signals.[2] Our evidence demonstrates that a language model can efficiently extract directional content at scale, alleviating this friction. We therefore offer an economic rationale for the emerging division of labor in which humans acquire and narrate soft information while machines improve the efficiency of information integration.

[2] Studies show that narratives regarding soft dimensions such as corporate culture, firm innovation, and strategic outlook are particularly informative (Bellstam et al., 2021; Lv, 2025; Li et al., 2026).

## 2. Related Literature

This study resides at the intersection of three literatures: information processing costs, expectations formation, and soft information in capital markets. Through the theoretical lens of "integration costs" and "quantification inefficiency," we synthesize and extend these disparate streams of literature. This section delineates our specific contributions to each.

### 2.1 Literature on disclosure processing costs

Our work belongs to the information economics literature examining the impact of information frictions on capital market outcomes. In their review, Blankespoor et al. (2020) unify this literature with a processing cost framework that decomposes the costs incurred by market participants into three sequential categories: the cost of becoming aware that a disclosure exists, the cost of acquiring the raw information signals within it, and the cost of integrating those multi-dimensional signals into a usable format for investing decisions. While this framework is central to understanding the informativeness of disclosure, cleanly isolating the specific costs—particularly *integration costs*—has remained an elusive empirical challenge.

In most archival settings, awareness, acquisition, and integration costs naturally co-vary because they are likely driven by the same features of a disclosure. When researchers examine cross-sectional variation in disclosure complexity (e.g., Gomez, 2024; Griffin et al., 2025), the documented differences in equity price adjustments therefore represent the combined effect of multiple frictions rather than a distinct consequence of any single cost. For instance, a dense, unreadable footnote section simultaneously raises the cost of locating and extracting specific contractual or tax line items—an acquisition cost—and the cognitive burden of synthesizing those line items into a formal asset valuation model—an integration cost. Therefore, prior evidence based on such proxies provides only indirect, suggestive support for integration costs, leaving the core empirical challenge unresolved.

To address this covariance problem, one stream of the literature has turned to controlled laboratory experiments (e.g., Elliott et al., 2022). By directly observing which participants acquire the disclosures, experimentalists can hold awareness and acquisition costs constant at zero, ensuring that any documented differences in forecast errors arise entirely from variation in integration costs. This method offers advantages in causal identification but is often limited in external validity and sample scale, and the degree to which findings generalize to sophisticated market participants such as financial analysts remains unclear.

Our within-report research design offers a novel solution to this identification problem. Using large-scale archival data, we create a natural laboratory: the textual narratives and numerical forecasts are issued simultaneously within a single sell-side analyst report. This allows us to hold the underlying information inputs and acquisition costs constant across the soft and hard outputs. That is, at the stage of integrating textual information into the summarizing numerical forecast, the acquisition costs of the raw qualitative insights embedded in the analyst's own written text are zero by construction. Thus, any systematic, predictable failure to translate the directional implications of the narrative into the numerical forecast can be explicitly attributed to integration cost. This design provides direct, real-world evidence on integration costs in Blankespoor et al.'s (2020) framework.

### 2.2 Literature on expectations formation and forecast efficiency

Our study also contributes to the literature on how professional forecasters, such as financial analysts, form expectations and why their forecasts systematically deviate from full-information rational expectations (FIRE). The standard empirical framework in this literature, established by Coibion and Gorodnichenko (2015, hereafter CG), tests FIRE by regressing *ex* *post* forecast errors on *ex ante* forecast revisions. Under FIRE, *ex ante* revisions should not predict *ex post* errors; therefore, any systematic relation between revisions and errors constitutes a rejection of FIRE.

Applying this test to consensus (average) forecasts, CG document that consensus revisions robustly predict consensus errors. They clarify that this consensus-level rejection of FIRE reflects *information frictions* while remaining consistent with *rational expectations*. In noisy-information models (Woodford, 2003; Sims, 2003), forecasters who observe noisy signals rationally underweight new information through Bayesian updating, causing average revisions to correlate with average errors. As CG (p. 2652) note, "the predictability of the average *ex post* forecast errors across agents using *ex ante* forecast revisions is an emergent property of the aggregation across individuals, not a property of the individual forecasts."[3]

In contrast, explaining an individual-level rejection of FIRE is far more restrictive. Under rational expectations, an individual forecast error must be orthogonal to every piece of information contained in that forecaster's information set. Because a forecast revision is part of this information set, any predictability of a forecaster's error using her own revision can only be attributed to *nonrational expectations*. This is why the literature documenting individual-level forecast predictability has turned to behavioral mechanisms—such as sticky expectations or diagnostic expectations (Bouchaud et al., 2019; Bordalo et al., 2020)—rather than to information frictions.

We argue that the CG framework is restrictive in a subtle but consequential way: it implicitly treats the numerical revision as a sufficient statistic for the forecaster's entire information set. This is valid only if quantifying qualitative signals is costless. When quantification is costly, a forecaster's information set contains not only the qualitative signals she has already quantified into her revision but also the qualitative signals she has acquired yet left unquantified—and a scalar revision cannot summarize the latter. By expanding the observable information set to include the directional content of the forecaster's own narrative (measured as textual tone), we show that individual forecasts systematically underreact to the forecaster's own acquired qualitative signals.

---

[3] Consensus-level predictability in the CG framework reflects the information awareness and acquisition costs formalized by Blankespoor et al. (2020), as noisy signals stem from individual analysts' inability to observe all signals available to the market (Woodford, 2003; de Silva and Thesmar, 2024). This information friction provides a theoretical explanation for early findings in accounting and finance that analyst consensus forecasts underreact to public signals such as past stock returns and earnings (e.g., Abarbanell, 1991; Abarbanell and Bernard, 1992; see Bradshaw, 2011 for a review).

Critically, this underreaction intensifies with our proxies for quantification costs—in vaguer reports, under higher environmental uncertainty, and for busier analysts—which locates the friction at the integration stage rather than during awareness or acquisition. We therefore provide an information-friction explanation for individual-level forecast bias, a mechanism that the standard CG framework cannot capture.[4]

### 2.3 Literature on soft information and textual analysis

Finally, our research extends the literature on the economic role of qualitative text (soft information) versus quantitative data (hard information) in corporate communication and capital markets. The existing textual analysis literature evaluates this soft-hard trade-off almost exclusively from the perspective of the *information user*: hard numerical metrics are precise, comparable, and costless to process, while soft text provides the rich narrative *context* required to correctly interpret, weight, and synthesize those numbers (Liberti and Petersen, 2019; Blankespoor et al., 2020). This framing explains why capital markets exhibit price reactions to the text of 10-K filings, conference calls, and financial press articles (see reviews by Li, 2010a; Loughran and McDonald, 2016).

[4] A nascent branch of this literature has begun to use natural language processing to measure forecasters' beliefs more richly. These studies, however, focus on text that forecasters either explicitly cite to justify their forecasts or have already incorporated into their quantitative models (e.g., Bastianello et al., 2025; Ke, 2025). By measuring only the quantified qualitative signals, they remain within the CG paradigm and continue to attribute residual individual-level predictability to nonrational expectations.

However, this framework fails to explain a persistent empirical puzzle: why textual information contains incremental value-relevant content beyond contemporaneous numerical indicators. This phenomenon is observed widely across corporate disclosures (Li, 2010b; Loughran and McDonald, 2011), analyst reports (Huang et al., 2014), news articles (Bybee et al., 2024), and central bank communications (Aruoba and Drechsel, 2026). If qualitative disclosure functions purely as context, it will only interact with hard numbers to trigger price reactions.

Our paper provides an explanation for this phenomenon by shifting the focus to the *information producer* (the analyst) and by distinguishing between "content" and "context" within high-dimensional soft information. We define *content* as standalone, value-relevant information that shifts expected value. Context, conversely, refers to interpretive information that guides decision making only through its interaction with existing content (e.g., Bochkay and Joos, 2021; Siano, 2025), such as by reducing the conditional variance of expected value.

Hardening soft information—quantifying qualitative text—reduces dimensionality, compressing high-dimensional text into a low-dimensional scalar value. Crucially, this dimensionality reduction affects content and context differently: while context is inevitably lost during quantification, primary content (captured by the scalar) is lost *if and only if* the information producer faces significant quantification costs. Because human analysts encounter binding cognitive and time constraints when mapping abstract real-world observations into precise numbers, they rationally choose to leave some value-relevant content unquantified and embedded within textual narratives. Therefore, quantification costs give rise to the incremental content observed in soft information.

Our empirical evidence directly validates this producer-side mechanism: an analyst's textual tone predicts her own numerical forecast errors, and this predictive power is two to nearly four times as strong in vaguer, harder-to-quantify reports. This finding establishes that text remains uniquely informative not because it is inherently superior to data, but because producers rely on narrative as an essential, cost-effective communication mechanism when faced with integration constraints. Our framework thus implies that the ultimate form in which information is disseminated to the market is governed by the relative processing costs of information producers versus users. This insight enriches our understanding of the soft-hard information trade-off.

## 3. Theoretical Framework and Hypotheses

To formally analyze the economic consequences of quantification friction, we develop a parsimonious model of how a rational but capacity-constrained analyst integrates qualitative and quantitative signals into a numerical forecast. The model's primary objective is to derive clear, testable hypotheses about how the quantification cost leads to predictable inefficiencies in the analyst's numerical forecasts. We anchor the model in the rational inattention and noisy information literature (Sims, 2003; Woodford, 2003), adapting it to a setting where the analyst must choose *whether* and *when* to transform vague qualitative knowledge into a precise number.

### 3.1 Model setup

#### *3.1.1 Decomposing earnings into quantitative and qualitative expectations*

Following de Silva and Thesmar (2024), we decompose earnings realizations and forecasts into distinct components. Consider an analyst who aims to forecast a firm's future earnings, $\pi$. The analyst acquires two distinct types of signals: a set of quantitative signals, $P$, with well-defined payoffs and probabilities (e.g., past earnings, stock returns, macroeconomic indicators), and a set of qualitative signals, $V$, which are valuable but vague, with ill-defined payoffs or probabilities (e.g., assessments of management strategy, competitive dynamics, or product pipeline quality).[5]

We decompose the earnings realization into components attributable to these information sets:

$$\pi = p + v + \varepsilon, \quad (1a)$$

where,

- $p = E(\pi|P)$ is the quantitative expectation—the rational expectation of earnings conditional only on the quantitative signals. This represents the forecast the analyst would make if she ignored all qualitative signals.
- $v = E(\pi|P,V) - E(\pi|P)$ is the qualitative expectation—the *incremental* shift in the earnings expectation when the analyst also conditions on the qualitative signals. It captures the value-relevant *content* of the qualitative signals. This is because if qualitative signals are redundant (i.e., $V \subseteq P$) or irrelevant to $\pi$, then $v = 0$; if qualitative signals only provide *context* for interpreting quantitative signals, which reduces conditional variance without shifting the conditional mean (i.e., $Var(\pi|P,V) < Var(\pi|P)$ but $E(\pi|P,V) = E(\pi|P)$), then $v$ remains zero. A positive (negative) $v$ implies the qualitative expectation is optimistic (pessimistic) relative to the quantitative baseline.
- $\varepsilon = \pi - E(\pi|P,V)$ is an unpredictable residual, orthogonal to the analyst's total information set, i.e., $E(\varepsilon|p,v) = 0$.[6]

[5] Qualitative signals are common in analyst information processing (Bradshaw, 2009). Analysts typically acquire such qualitative signals through in-person communication, including conference calls (Huang et al., 2018; Dzielinski et al., 2021), private meetings (Soltes, 2014), site visits (Cheng et al., 2016), and broker conferences (Green et al., 2014). Analysts also extract qualitative insights from corporate textual disclosures (Lehavy et al., 2011), news articles (Bradshaw et al., 2021), and their own prior industry experience (Bradley et al., 2017).

[6] Because $P$ and $V$ represent signals acquired by the analyst, the residual reflects two distinct sources of unpredictability. The first source is unacquired signals, which exist and may be known by others but are not acquired by the analyst, perhaps due to awareness and acquisition costs. In the framework of Woodford (2003), such omission renders $P$ and $V$ noisy information relative to all existing signals. The second source is earnings

Crucially, the qualitative expectation $v$ is a latent variable. While the analyst perceives the directional implication of the qualitative signals (i.e., the sign of $v$), she cannot directly observe its precise cardinal value. That is, the analyst knows that the firm's prospects are better or worse because of the qualitative insights, but she lacks a precise numerical estimate of how much better or worse.[7]

### *3.1.2 The quantification friction in expectations formation*

Producing a numerical forecast requires the analyst to translate imprecise qualitative signals into a precise number. This act of quantification is costly for a capacity-constrained human information processor. Specifically, the analyst must expend cognitive effort and research time to resolve the vagueness of qualitative signals and refine their implications for future earnings. Under the processing cost framework of Blankespoor et al. (2020), this quantification cost is an integration cost.

We model this friction by introducing a parameter, $\theta \in [0,1]$, which represents the fraction of the qualitative expectation ($v$) that remains *unquantified* due to the underlying quantification cost, $c \in [0, \infty]$. This parameter partitions the qualitative expectation into a quantified (integrated) component and an unquantified (unintegrated) component, modifying the earnings realization expression from equation (1a) to:

$$\pi = p + \underbrace{(1-\theta)v}_{\substack{quantified\\component}} + \underbrace{\theta v}_{\substack{unquantified\\component}} + \varepsilon, \tag{1b}$$

where $\theta = \Theta(c)$ is an increasing function of the quantification cost (i.e., $\Theta'(c) > 0$). When quantification is costless ($c = 0$), $\theta = 0$, indicating that the qualitative expectation is fully innovation beyond all existing signals, representing the "unknown unknowns" that even a fully informed analyst could not see.

[7] For example, a Morgan Stanley report on Starbucks (John Glass and team, February 18, 2009) noted that store-level labor cost reductions may take longer "as it requires a greater cultural change." Although the directional delay is clear, "cultural change" lacks a well-defined, quantifiable metric. Similarly, a Credit Suisse report on Pfizer (Catherine J. Arnold and team, November 1, 2011) stated that tofacitinib's profile "sets it up to be a blockbuster" conditional on approval. This signal implies a strong positive outlook but provides no precise numerical valuation for a "blockbuster" or specific probability estimate for the drug's approval.

integrated into the numerical forecast. Conversely, as the quantification cost approaches infinity ($c \to \infty$), $\theta \to 1$, capturing that the analyst leaves her qualitative expectation almost entirely unquantified.

Correspondingly, we express the analyst's numerical forecast, $F$, as the sum of three terms: an information term, a bias term, and a cognitive noise term:

$$F = \underbrace{p + (1-\theta)v}_{information} + \underbrace{b}_{bias} + \underbrace{\eta}_{noise}, \tag{2}$$

where

- $p + (1-\theta)v$ is the information term, which represents the rational expectation of earnings subject to the quantification friction;
- $b = E(F - \pi|P)$ represents systematic forecast bias predicted by quantitative signals; and
- $\eta = F - E(F|P, V)$ represents random cognitive noise that is orthogonal to all signals, i.e., $E(\eta|P, V) = 0$.[8]

Because a numerical forecast requires precision, the information term only includes the quantitative expectation $p$ and the quantified component of the qualitative expectation $(1-\theta)v$.

[8] Following de Silva and Thesmar (2024), we relax the rational expectations assumption and align the model with empirical observations by including bias ($b$) and cognitive noise ($\eta$). The bias term captures well-documented strategic and behavioral biases, for which we take no stance on the source. To cleanly isolate the quantification friction and maintain analytical tractability, we assume that these biases are conditional only on quantitative signals $P$. We later relax this assumption in Section 3.4 by allowing these biases to relate to qualitative signals $V$. The cognitive noise term captures the phenomenon where forecasters generate different forecasts based on identical signals (Kahneman et al., 2021). This is conceptually distinct from noisy information (Woodford, 2003), as captured by $\varepsilon$ in equation (1a), which arises from the information awareness and acquisition process.

### 3.2 Incomplete quantification

#### *3.2.1 Main hypothesis: Predictable forecast errors*

To capture the structural impact of this quantification friction on forecast efficiency, we define the contemporaneous forecast error as $F - \pi$. Subtracting the earnings realization expression in equation (1b) from the forecast expression in equation (2) yields:

$$F - \pi = -\theta v + b + \eta - \varepsilon. \tag{3}$$

Equation (3) is the heart of our first hypothesis. Under full-information rational expectations, forecast errors must be orthogonal to all signals in the analyst's information set (i.e., $E(F - \pi | P, V) = 0$). However, when information integration is subject to the quantification cost (i.e., $c > 0 \Rightarrow \theta > 0$), the contemporaneous forecast error becomes a direct, decreasing function of the qualitative expectation $v$ (i.e., $\frac{\partial E(F-\pi|P,V)}{\partial v} = -\theta < 0$).

Consequently, the analyst's qualitative expectation systematically predicts her own *ex post* forecast error. To see the economic intuition, consider an analyst who unearths favorable soft information about firm operations (e.g., a positive shift in corporate culture or a successful hiring of key talent) and forms a positive qualitative expectation ($v > 0$). Because a fraction of this qualitative insight ($\theta v$) remains unquantified, her numerical forecast underweights this positive news. The result is a forecast that is systematically too low relative to realized earnings, $\pi$, generating a negative forecast error (i.e., $F - \pi = -\theta v < 0$, abstracting from bias, noise, and shocks).

Empirically, we use the machine-extracted overall textual tone of the analyst's report as our operational proxy for the latent qualitative expectation $v$. As long as the analyst describes the qualitative signals she has acquired in the report text, the overall tone of this text captures the directional implication of these signals and thus indicates the latent qualitative expectation.[9] This leads to our first hypothesis:

***H1 (Incomplete Quantification):*** *If quantification costs prevent an analyst from fully integrating qualitative signals into numerical estimates, the overall textual tone of the analyst's report will negatively predict her contemporaneous forecast error.*

### *3.2.2 Cross-sectional variation: Quantification difficulty and linguistic vagueness*

Our theoretical framework implies that the degree of underreaction, $\theta$, is increasing in the underlying quantification cost, $c$. To validate that the forecast inefficiency predicted by H1 is driven by this quantification friction rather than by alternative behavioral or strategic channels, we exploit cross-sectional proxies for the quantification cost $c$.

The first proxy is linguistic vagueness. When report text is more qualitative (i.e., contains fewer numbers) and employs more hedging language, the quantification cost $c$ escalates. Such text reflects a less precise foundation of qualitative signals, making the mapping of these signals into a line-item earnings estimate more effortful. As the linguistic vagueness of a report increases, the quantification cost $c$ rises, inflating $\theta$. From equation (3), a larger $\theta$ amplifies the negative relation between report tone and the contemporaneous forecast error, yielding our next testable prediction:

***H1a (Quantification Difficulty):*** *The predictive power of report tone for contemporaneous forecast errors is stronger in reports characterized by a higher degree of linguistic vagueness.*

[9] Report text provides a robust proxy for otherwise unobservable qualitative signals. We do not assume that report text captures qualitative expectation ($v$) completely or exclusively. If report text contains only context for interpreting the quantitative expectation, irrelevant information, or noise stemming from qualitative signals, tone would not correlate with $v$. This measurement error would attenuate our estimated coefficients, shifting our tests toward the null of informational efficiency and ensuring a robust test for inefficiency (Bradshaw 2011; de Silva and Thesmar 2024). Alternatively, if report text reflects irrational sentiment rather than fundamental information (Tetlock, 2007; Garcia, 2013; Goetzmann et al., 2024), tone would correlate with bias ($b$) rather than qualitative expectation ($v$). This implies an overreaction (Bastianello et al., 2025; Ke, 2025), running counter to the underreaction predicted by our model.

#### *3.2.3 Cross-sectional variation: Processing capacity constraints and analyst busyness*

Beyond the characteristics of qualitative signals themselves, quantification costs are driven by the exogenous supply of an analyst's cognitive processing capacity. In the spirit of rational inattention frameworks (Sims, 2003), an analyst allocates limited mental bandwidth across her entire coverage portfolio. When an analyst's coverage expands or when multiple firms in her portfolio issue concurrent disclosures, numerically modeling all qualitative signals becomes less economically feasible. Consequently, busier analysts will leave a greater share of their qualitative insights unquantified (i.e., $\theta$ increases), relying more heavily on narrative disclosure to communicate these insights to clients. This yields our next hypothesis:

***H1b (Capacity Constraints):*** *The predictive power of report tone for contemporaneous forecast errors is stronger when the issuing analyst is busier.*

### 3.3 Delayed quantification

#### *3.3.1 Main hypothesis: Predictable subsequent revisions*

Next, we extend the framework to a dynamic horizon to capture how the quantification friction operates over time. While narrative communication serves as a repository for acquired but unquantified content ($\theta v$), this component does not remain permanently unintegrated. Instead, as the forecast target event approaches, the arrival of additional public data, the provision of clearer firm disclosures, and the completion of sequential research tasks lower the downstream integration barrier.

Following costly learning models (Cukierman, 1980; Dugast and Foucault, 2018; Banerjee et al., 2020), we assume that between time $t$ and $t + 1$, the analyst accumulates incremental knowledge that renders the previously unquantified qualitative component ($\theta_t v_t$) quantifiable, yet potentially noisy:

$$e_{t+1} = \theta_t v_t + \xi_{t+1}, \tag{4}$$

where $\xi_{t+1}$ represents a noise term orthogonal to the underlying qualitative signals, i.e., $E(\xi_{t+1}|\theta_t v_t) = 0$. Upon observing this updated signal, the analyst executes a Bayesian update to revise her numerical forecast. Weighting the precision of the updated signal against her existing forecast, the revised forecast $F_{t+1}$ is expressed as:

$$F_{t+1} = \lambda e_{t+1} + (1-\lambda)F_t, \tag{5}$$

where $\lambda \in [0,1]$ represents the Kalman gain derived from the signal-to-noise ratio of the newly quantified signal.[10]

To express the subsequent forecast revision, we take the first-order difference of the forecast series as:

$$\Delta F_{t+1} = \lambda e_{t+1} - \lambda e_t + (1-\lambda)\Delta F_t. \tag{6a}$$

Taking the conditional expectation of this forecast revision with respect to the current and prior qualitative expectations ($v_t$, $v_{t-1}$) and the current forecast revision ($\Delta F_t$) yields:

$$E(\Delta F_{t+1}|v_t, v_{t-1}, \Delta F_t) = \lambda\theta_t v_t - \lambda\theta_{t-1}v_{t-1} + (1-\lambda)\Delta F_t. \tag{6b}$$

Equation (6b) provides the foundation for our empirical analysis of forecast dynamics. It shows that the direction and magnitude of the analyst's subsequent revision ($\Delta F_{t+1}$) are a positive function of her current qualitative expectation $v_t$ (i.e., $\frac{\partial E(\Delta F_{t+1}|v_t, v_{t-1}, \Delta F_t)}{\partial v_t} = \lambda\theta_t > 0$). The economic intuition is that while initial quantification costs caused the analyst to undershoot her qualitative insight at time $t$ ($\theta_t > 0$), the gradual easing of this friction ($\lambda > 0$) generates a predictable subsequent adjustment path: the analyst moves her numerical forecast step by step in the exact direction initially charted by her report text. This leads to our second main hypothesis:

[10] Specifically, $\lambda = \frac{\tau_e}{\tau_F + \tau_e}$, where $\tau_e$ and $\tau_F$ denote the precision (the inverse of variance) of the newly quantified signal and existing forecast, respectively. We distinguish between noisy signals ($\tau_e$ is low but positive) and vague signals (ill-defined, unmeasurable, and thus $\tau_e$ is effectively zero). At time $t$, qualitative signals remain vague and receive no weight ($\lambda = 0$). By time $t+1$, as precision improves through time-consuming research (Cukierman, 1980; Dugast and Foucault, 2018) or the resolution of uncertainty (Banerjee et al., 2020), these signals become quantifiable and receive positive weights ($\lambda > 0$). For analytical tractability, we treat the Kalman gain $\lambda$ as a structural constant representing the speed at which this precision improves subsequently.

***H2 (Delayed Quantification):*** *If quantification costs delay the integration of qualitative signals, the overall textual tone of the analyst's current report will positively predict her subsequent numerical revisions.*

### *3.3.2 Cross-sectional variation: Evidence requirements and environmental uncertainty*

Equation (6b) demonstrates that the magnitude of the subsequent forecast drift depends on the size of the initial unquantified component, $\theta_t$.

Under Knightian uncertainty or severe information asymmetry, mapping unstructured qualitative signals to concrete earnings metrics is costlier and requires more extensive evidence to construct reliable mathematical forecasting models. Consequently, analysts leave a larger fraction of value-relevant signals unquantified at time $t$ (i.e., $\theta_t$ increases), deferring quantification until subsequent information arrives to resolve the underlying uncertainty. This mechanism implies that the positive predictive link between current textual tone and subsequent forecast revisions should be amplified in more volatile or opaque environments. This leads to our third cross-sectional hypothesis:

***H2a (Evidence Requirement):*** *The predictive power of report tone for subsequent forecast revisions is stronger when the analyst operates under higher levels of environmental uncertainty.*

### *3.3.3 Cross-sectional variation: Processing capacity constraints and analyst busyness*

Similarly, processing capacity constraints prevent busy analysts from immediately integrating qualitative signals into their financial models. These resource-constrained analysts must defer a larger fraction of the quantification process (i.e., $\theta_t$ increases), resulting in a more pronounced delayed adjustment path. This leads to our final cross-sectional hypothesis:

***H2b (Capacity Constraints):*** *The predictive power of report tone for subsequent forecast revisions is stronger when the issuing analyst is busier.*

### 3.4 Delineation from alternative explanations

To isolate the quantification friction, our model establishes an orthogonality condition between the qualitative expectation ($v$) and strategic or behavioral biases. Relaxing this condition allows equation (3) to be interpreted alternatively as:

$$F - \pi = \underbrace{-\theta v}_{bias\ conditional\ on\ V} + \underbrace{b}_{bias\ conditional\ on\ P} + \eta - \varepsilon.$$

Under this interpretation, the underreaction to the qualitative expectation ($-\theta v$) could reflect strategic or behavioral biases related to qualitative signals rather than the friction driven by the quantification cost. We address three primary alternative explanations that could generate this pattern.

First, the underreaction may represent optimistic bias. The prior literature on sell-side analysts suggests that analysts manipulate outputs to curry favor with corporate executives or to stimulate brokerage business (e.g., Lim, 2001; Chen and Jiang, 2006). If our results were entirely an artifact of intentional optimistic bias—wherein analysts purposefully relegate adverse news to the less visible text section—the predictable errors would be driven exclusively by negative textual signals ($v < 0$). In contrast, our model predicts that the quantification friction ($\theta$) operates symmetrically across both positive and negative qualitative expectations.

Second, the underreaction may reflect institutional catering or reputational concerns. Analysts may intentionally keep public numerical outputs sticky to selectively offer an information edge to private institutional clients via text (Amiram et al., 2025; Bradshaw et al., 2025; Xiao and Zang, 2025) or to avoid reputational damage from seesaw revisions under uncertainty (Raedy et al., 2006). While such strategic underreaction varies with institutional catering incentives and reputational pressures, the quantification friction ($\theta$) is a structural feature of human processing limits. Consequently, our framework implies that quantification inefficiency will persist even in environments where institutional catering incentives and reputational pressures are minimized or absent.

Third, the underreaction may reflect behavioral biases such as conservatism or anchoring (Edwards, 1968; Tversky and Kahneman, 1974). However, naive, psychology-driven forecasting errors typically diminish as professional information processors learn from past errors (Mikhail et al., 2003). Because we model the degree of underreaction ($\theta$) as a friction rooted in the quantification process, it should remain robust across both low- and high-experience analyst segments.

The empirical analyses detailed in Section 6 provide direct tests to distinguish our mechanism from these alternative explanations.

## 4. Sample and Variables

### 4.1 Sample selection

We construct the sample by matching analyst reports from Investext to the corresponding analyst forecasts in I/B/E/S. Table 1 outlines the sample selection process. We begin with 635,041 analyst reports covering S&P 500 firms from 1995 to 2012. Matching these reports to I/B/E/S forecasts by analyst and broker names yields 472,269 matched reports. To ensure that each report contains both textual content and a contemporaneous numerical forecast, we follow Huang et al. (2014) and match Investext report dates to the [-2, +2] window around the I/B/E/S announcement dates of the analyst's earnings forecasts, target prices, and stock recommendations for the same firm.[11] The final sample consists of 241,045 reports with at least one numerical forecast, comprising 209,723 earnings forecasts, 88,849 target prices, and 24,986 stock recommendations.[12]

[11] This narrow window ensures that the textual and numerical outputs we link are produced as part of the same report event.

[12] Earnings forecast, target price, and stock recommendation revisions represent 44%, 19%, and 5% of the matched reports, respectively. These proportions are consistent with Huang et al. (2014).

*[Insert Table 1 here]*

## 4.2 Variable measurement

To proxy for the latent qualitative expectation $v$, we measure the overall textual tone of each analyst report (i.e., $Tone$) in three steps. First, we remove boilerplate brokerage disclosures following Huang et al. (2014). Second, we apply the FinBERT classifier of Huang et al. (2023) to classify each sentence as positive, neutral, or negative.[13] Third, we calculate $Tone$ as the number of positive sentences minus the number of negative sentences, scaled by the total number of sentences in the report.

Because qualitative expectations lack a precise cardinal magnitude, $Tone$ does not measure $v$ directly; it captures the overall directional orientation of the report text instead. This is consistent with our theoretical construct in Section 3.1, under which the analyst perceives the sign of her qualitative expectation without observing its precise value. By collapsing high-dimensional text into a single directional score, the FinBERT classification provides a parsimonious, quantifiable proxy for the underlying qualitative expectation.

To capture cross-sectional variation in quantification costs, $c$, we construct two measures of linguistic vagueness. The first, $TextOnly\%$, is the percentage of report sentences that contain no numerical information. Following Huang et al. (2014), we identify a sentence as number-free when it contains neither a \$ nor a % sign, since numerical content in analyst reports is typically expressed in dollar amounts or percentage changes. A sentence without numbers is inherently vaguer than one containing numbers. Two sentences from a Starbucks report illustrate the difference:

[13] Huang et al. (2023) developed FinBERT by fine-tuning Google's BERT on a large corpus of corporate disclosures and analyst reports. They show that FinBERT captures the contextual information in analyst report text and outperforms dictionary-based methods and other machine-learning algorithms. BERT is a neural network-based large language model built on a transformer architecture and trained over 3.3 billion words, achieving human-level performance across various language processing tasks (e.g., Michaelov et al., 2021).

(1) While still early, SBUX has a newfound willingness to take necessary measures to cut costs, close stores and repair margins as it moves from growth to maturity.

(2) Recent investor meetings we hosted with management increased our confidence in SBUX's ability to obtain $500M in cost savings in FY09 alone, but caution investors not to expect an overnight revival in top-line results.[14]

The first sentence's qualitative description of cost-cutting is inherently vaguer than the second one's explicit $500M savings estimate.

The second measure, $Hedge\%$, is the percentage of sentences containing hedging words—terms that explicitly acknowledge imprecision. Based on Lakoff (1973), Channell (1994), and Loughran and McDonald (2011), we compile a hedging wordlist across five categories (see the Appendix for the full list):

- subjective beliefs (e.g., "think," "believe," and "suggest");
- vague probability (e.g., "seem," "appear," "may," "might," "maybe," "perhaps," "possibly," "likely," and "probably");
- vague quantity and frequency (e.g., "around," "roughly," "some," "many," "sometimes," and "often");
- vague extent (e.g., "kind of," "fairly," "almost," "so to speak," and "to some extent"); and
- vague manner (e.g., "usually," "somehow," and "in a sense").

A sentence with such hedge words signals lower precision of the underlying information than unhedged ones. In a Pfizer report, for example, the analyst wrote:

> Pfizer may not even be able to achieve $2.00 in 2012 because revenue will fall materially short of management's $70B target.[15]

[14] John Glass and team, Morgan Stanley, February 18, 2009.
[15] David Risinger and team, Morgan Stanley, September 10, 2009.

Here, “may” indicates that the analyst lacks a sufficient basis to warrant the $2.00 forecast, revealing her low confidence in that numerical estimate. We classify a report as vaguer when $TextOnly\%$ is higher, or separately, when $Hedge\%$ is higher.

We examine three numerical outputs: earnings forecasts, target prices, and stock recommendations. We measure earnings forecast error ($FError$) as the earnings forecast minus the actual earnings; we measure the subsequent earnings forecast revision ($FRev_{t+1}$) as the analyst’s subsequent earnings forecast for the same firm-fiscal year minus her contemporaneous forecast. Because not all contemporaneous forecasts are followed by such a revision, the forecast revision sample is slightly smaller than the forecast error sample. Control variables for earnings forecast characteristics include forecast horizon ($Horizon$) and boldness ($Bold$, an indicator for forecasts deviating significantly from the prevailing forecast consensus). We define subsequent revisions for target prices ($TPRev_{t+1}$) and stock recommendations in the same way ($RRev_{t+1}$) as the difference between subsequent and contemporaneous values. To mitigate firm heterogeneity and outlier effects, the continuous numerical variables (i.e., $FError$, $FRev_{t+1}$, $TPRev_{t+1}$) are scaled by the firm’s stock price 50 days prior to the contemporaneous forecast (or subsequent revision) and winsorized at the 1st and 99th percentiles.

Following the analyst-forecast literature (e.g., Cooper et al., 2001; Clement and Tse, 2005), our regressions include analyst, broker, and firm controls. Analyst characteristics comprise general experience ($AnaGenExp$), firm-specific experience ($AnaFirmExp$), star status ($AnaAllStar$), the leader-follower ratio ($AnaLFR$), firm coverage ($AnaNFirm$), and forecast frequency ($AnaNFore$). Broker characteristics comprise brokerage house size ($BHSize$) and investment bank status ($BHIB$). Firm characteristics comprise firm size ($FirmSize$), book-to-market ratio ($FirmBM$), institutional ownership ($FirmIO$), and the information environment, proxied by analyst coverage ($FirmNAna$) and the number of forecasts issued for the firm ($FirmNFore$). We also control for recent firm news, measured by cumulative abnormal returns prior to the current report and the subsequent report ($PriorCAR_{t \text{ or } t+1}$, respectively). Data come from I/B/E/S, CRSP, Compustat, SDC, Thomson Financial 13F, and Capital IQ's Key Developments. The Appendix provides detailed variable definitions.

### 4.3 Descriptive statistics

Table 2 presents descriptive statistics. Mean $Tone$ is 0.14, and about one-quarter of reports have a net negative tone, indicating that analysts' qualitative views skew optimistic.[16] The average report contains 48 sentences—a substantial volume of qualitative content. Of these, 57% contain no numerical information ($TextOnly\%$), and 38% contain at least one hedging word ($Hedge\%$), implying that quantification costs are non-negligible for a large share of report content.

*[Insert Table 2 here]*

We sort reports into quintiles of contemporaneous $Tone$, with Q1 the most pessimistic and Q5 the most optimistic. Figure 1 plots mean forecast errors by quintile, providing preliminary visual evidence consistent with quantification inefficiency. The most pessimistic $Tone$ quintile (Q1) exhibits the highest, positive mean $FError$, suggesting that when analysts write most pessimistically, their earnings forecasts are too high relative to realizations. The most optimistic $Tone$ quintile (Q5) exhibits the lowest, negative mean $FError$, indicating that when analysts write most optimistically, their earnings forecasts are too low relative to realizations. Mean $FError$ declines monotonically from Q1 to Q5. This pattern persists for forecasts issued up to four subsequent reports (i.e., $t + 1$ through $t + 4$), though the Q1–Q5 spread narrows as uncertainty resolves over time. These results suggest that textual tone predicts the direction of bias in an analyst's subsequent numerical forecasts.

[16] While the level of $Tone$ is positively skewed, untabulated statistics show that $\Delta Tone$ (i.e., $Tone$ of the current report minus $Tone$ of the prior report for the same firm) is distributed symmetrically around zero, suggesting that analysts are equally likely to revise their qualitative outlook in either direction.

*[Insert Figure 1 here]*

# 5. Empirical Results

## 5.1 Incomplete quantification

H1 predicts a negative relation between report tone ($Tone$) and the contemporaneous forecast error ($FError$, measured as the earnings forecast minus actual earnings). We test this prediction by estimating the following OLS regression, which mirrors the structural relation in equation (3):

$$FError = \alpha + \beta_1 Tone + \Sigma \Upsilon_i \text{Controls}_i + \mu. \quad (7)$$

We predict a negative coefficient on $Tone$. All specifications include the control variables described in Section 4.2, with progressively stringent sets of fixed effects.

We first estimate equation (7) with year fixed effects, which control for common time trends in forecast errors. Column 1 of Table 3 reports the results. Consistent with H1, the coefficient on $Tone$ is negative and significant at the 1% level. In terms of economic magnitude, a one-standard-deviation increase in $Tone$ (0.31) is associated with a decrease in $FError$ of approximately 6% of its standard deviation. The results suggest that analysts' contemporaneous forecasts thus do not fully incorporate the qualitative signals conveyed in their own report text.

*[Insert Table 3 here]*

We next add analyst-year fixed effects, which absorb analyst characteristics such as experience, portfolio size, and persistent forecasting style. This specification asks whether, within an analyst's coverage portfolio in a given year, $Tone$ is negatively related to $FError$ across the firms. Column 2 shows that the coefficient on $Tone$ remains negative and significant at the 1% level: when an analyst writes optimistically (pessimistically) about a firm, her contemporaneous forecast for that firm tends to be too low (high) relative to her forecasts for other covered firms. This pattern confirms that the predictable forecast error is not attributable to analyst characteristics that vary across analysts or years.

Finally, we include analyst-firm and year fixed effects. Analyst-firm fixed effects absorb all time-invariant firm characteristics (e.g., size, institutional ownership, analyst coverage) and analyst-firm characteristics (e.g., familiarity, the firm's importance to the analyst), so identification comes from variation across forecasts issued by the same analyst for the same firm over time. Column 3 shows that the coefficient on $Tone$ remains negative and significant at the 1% level, indicating that our findings are not driven by analyst-specific forecasting habits toward particular firms. In this more stringent specification, the economic magnitude is attenuated but remains economically meaningful: a one-standard-deviation increase in $Tone$ still lowers $FError$ by about 2% of its standard deviation.

Taken together, the three specifications consistently support H1: analysts do not fully integrate qualitative signals in their own report text into contemporaneous earnings forecasts, leaving forecast errors predictably related to textual tone.[17]

### 5.2 Cross-sectional variation based on linguistic vagueness

H1a predicts that incomplete quantification is more pronounced in linguistically vaguer reports. To test this prediction, we augment equation (7) with an indicator variable for vaguer reports ($Vagueness$) and its interaction with tone ($Tone \times Vagueness$):

[17] We conduct several robustness tests (see results in the Internet Appendix). First, we control for additional signals by replacing $PriorCAR$ with a longer window return ($2MPriorCAR$), adding prior annual earnings surprises ($ESurp$), and incorporating forecast revisions ($FRev_t$) to reconcile our findings with the Coibion and Gorodnichenko (2015) framework. Second, we exclude prompt reports (i.e., those issued concurrently with earnings announcements or management forecasts) to mitigate the influence of confounding information events. Third, to isolate tone from personal writing style, we use both the change in $Tone$ relative to the analyst's prior report and an analyst-firm-demeaned $Tone$ measures. Across all tests, our results remain qualitatively similar to our main findings, confirming that our evidence is robust to alternative information controls and tone measures.

$$FError = \alpha + \beta_1 Tone + \beta_2 Tone \times Vagueness + \beta_3 Vagueness + \Sigma \Upsilon_i \text{Controls}_i + \mu, \quad (8)$$

where $Vagueness$ equals one when a measure of linguistic vagueness from Section 4.2—either the percentage of number-free sentences ($TextOnly$%) or the percentage of hedged sentences ($Hedge$%)—exceeds its median across the analyst's own reports, and zero otherwise. We predict a negative coefficient on the interaction term $Tone \times Vagueness$.

Table 4 reports the results. In column 1, where $Vagueness$ is based on the absence of numbers, the estimated coefficient on $Tone \times Vagueness$ is negative and significant at the 5% level. This result suggests that the qualitative content of number-avoiding reports is less fully incorporated into the analyst's contemporaneous forecast. In column 2, where $Vagueness$ is based on hedging words, the estimated coefficient on $Tone \times Vagueness$ is again negative and significant at the 1% level, suggesting that report content conveyed with explicit acknowledgments of imprecision is likewise less fully quantified. In terms of economic magnitudes, the predictive power of tone for forecast errors is two to nearly four times as strong in vaguer reports than in other reports, as measured by the relative size of the estimated coefficients on $Tone \times Vagueness$ and $Tone$.

*[Insert Table 4 here]*

Overall, the cross-sectional evidence supports H1a: underreaction to textual tone concentrates in reports that are harder to quantify, indicating that the observed quantification inefficiency is driven by the difficulty of quantifying vague narrative signals.

### 5.3 Delayed quantification

H2 predicts a positive relation between the tone of the current report ($Tone$) and the analyst's subsequent revisions of her numerical forecasts—earnings forecasts ($FRev_{t+1}$), target prices ($TPRev_{t+1}$), and stock recommendations ($RRev_{t+1}$). We test this prediction by estimating the following OLS regression, which mirrors the structural relation in equation (6b), and we expect a positive coefficient on $Tone$:

$$FRev_{t+1}\ (or\ TPRev_{t+1}, RRev_{t+1}) = \alpha + \beta_1 Tone + \Sigma\Upsilon_i \text{Controls}_i + \mu. \qquad (9)$$

All specifications control for the contemporaneous revision in the corresponding numerical forecasts (i.e., $FRev_t$, $TPRev_t$, or $RRev_t$), the control variables described in Section 4.2, and the same fixed effects as in equation (7); $Horizon$ and $Bold$, which are specific to earnings forecasts, are excluded from the target-price and recommendation regressions.

Table 5 reports the results. Panel A presents the earnings forecast revisions. The estimated coefficients on $Tone$ are positive and significant at the 1% level in all three fixed-effect specifications, consistent with H2. In terms of economic magnitude, a one-standard-deviation increase in $Tone$ (0.31) is associated with an increase in $FRev_{t+1}$ of 5% to 6% of its standard deviation.[18]

*[Insert Table 5 here]*

Panel B presents the target price revisions.[19] The estimated coefficients on $Tone$ remain positive and significant at the 1% level across the three specifications, indicating that delayed quantification extends beyond earnings forecasts.

Panel C presents the stock recommendation revisions. Because stock recommendations have only three or five discrete levels (i.e., Strong Buy/Buy, Hold, and Sell/Strong Sell), revisions are mechanically bounded by the current recommendation level; we therefore estimate equation (9) separately for current Buy, Hold, and Sell reports. The estimated coefficient on $Tone$ is positive and significant at the 1% level for current Buy and Hold reports (columns 1–6), consistent with H2. It is not robust for current Sell reports (columns 7–9), likely reflecting the small number of Sell reports in the sample.

---

[18] We conduct the same robustness tests for equation (9) as described in footnote 17. Additionally, we test whether the $Tone$ of an analyst's current report continues to predict the earnings forecast revisions of multiple subsequent reports and find that $Tone$ predicts $FRev_{t+2}$ and $FRev_{t+3}$. See the Internet Appendix for the results.
[19] The analysis of target price revisions is based on the sample excluding analyst reports from Morgan Stanley, which provides range forecasts of target prices (Joos et al., 2016; Bochkay and Joos, 2021).

Taken together, the evidence across all three numerical outputs supports H2 that analysts do not immediately quantify the qualitative signals in their report text; instead, the directional content embedded in text flows into subsequent numerical revisions as new evidence arrives and quantification friction eases.

### 5.4 Cross-sectional variation based on environmental uncertainty

H2a predicts that environmental uncertainty amplifies quantification delays. Because uncertainty is a broad and amorphous concept for which there is "no perfect measure but instead a broad range of proxies" (Bloom, 2014, p.154), we employ five sets of proxies suggested by prior literature: (1) long forecast horizons, (2) high macroeconomic and market volatility, (3) high firm-level volatility, (4) significant corporate changes, and (5) poor information environments.

In the first test, following Raedy et al. (2006), we partition the sample into four groups by forecast horizon—the number of days between the forecast and the earnings announcement. Because information accumulates over time, forecasts issued further in advance face higher uncertainty; H2a therefore predicts that the coefficient on $Tone$ in equation (9) increases with the forecast horizon.

Table 6, Panel A reports the results. Both the magnitude and the significance of the estimated coefficients on $Tone$ increase monotonically with the forecast horizon, from less than one quarter ahead (0–89 days, column 1) to three quarters ahead (270–360 days, column 4), consistent with H2a. Notably, $Tone$ does not predict $FRev_{t+1}$ within one quarter, suggesting that when little uncertainty remains, there is little unquantified content left to flow into future revisions.

*[Insert Table 6 here]*

For the remaining uncertainty tests, we estimate the following equation:

$$FRev_{t+1} = \alpha + \beta_1 Tone + \beta_2 Tone \times Uncertainty + \beta_3 Uncertainty + \Sigma \Upsilon_i \text{Controls}_i + \mu, \quad (10)$$

where $Uncertainty$ equals one when the respective uncertainty measure exceeds its sample median, and zero otherwise; we predict a positive coefficient on the interaction $Tone \times Uncertainty$. Panel B follows Loh and Stulz (2018) and Bochkay and Joos (2021) and measures macroeconomic and market uncertainty using the uncertainty index developed by Jurado et al. (2015) (column 1), recession periods identified by the National Bureau of Economic Research (column 2), and the VIX (column 3).[20] Panel C follows Zhang (2006a, 2006b) and uses firm-level volatility measured as five-year earnings volatility (column 1) and one-year stock-return volatility (column 2). Panel D follows Boulland et al. (2023) and identifies significant corporate changes including mergers and acquisitions (column 1), changes in the number of business segments (column 2), and CEO or CFO turnover (column 3). Panel E follows Lang and Lundholm (1996) and Zhang (2006a, 2006b) and measures $Uncertainty$ as poor information environments, using small firm size (column 1), low analyst coverage (column 2), and high analyst forecast dispersion (column 3). Across Panels B–E, the estimated coefficients on $Tone \times Uncertainty$ are positive and significant at least at the 5% level, consistent with H2a.

Overall, the cross-sectional evidence supports H2a. That is, when uncertainty is high, analysts require more evidence before committing qualitative signals to numbers, hence, textual tone has greater predictive power for subsequent forecast revisions.

### 5.5 Cross-sectional variation based on analyst busyness

H1b and H2b predict that quantification inefficiency increases with processing-capacity constraints: that is, tone should have greater predictive power for both contemporaneous forecast errors and subsequent forecast revisions when the analyst is busier. We test these predictions with two proxies for analyst $Busyness$. First, following Harford et al. (2019), $Busyness$ equals one when the number of firms in the analyst's coverage portfolio exceeds the sample median, capturing limited time per covered firm; and zero otherwise. Second, following Driskill et al. (2020), $Busyness$ equals one for reports issued on "busy days," defined as days on which more than one covered firm releases an earnings announcement or management forecast within the [-1, +1] window, capturing concurrent demands on the analyst's attention; and zero otherwise.[21]

[20] The Jurado-Ludvigson-Ng index captures the conditional volatility of the unforecastable component of macroeconomic indicators. By contrast, VIX captures factors of conditional volatility other than macroeconomic uncertainty such as changes in investor sentiment or risk preferences (Jurado et al., 2015).

To test the two hypotheses, we modify equations (7) and (9) by including the indicator $Busyness$ and the interaction term $Tone \times Busyness$:

$$FError = \alpha + \beta_1 Tone + \beta_2 Tone \times Busyness + \beta_3 Busyness + \Sigma \Upsilon_i \text{Controls}_i + \mu; \quad (11)$$

$$FRev_{t+1} = \alpha + \beta_1 Tone + \beta_2 Tone \times Busyness + \beta_3 Busyness + \Sigma \Upsilon_i \text{Controls}_i + \mu. \quad (12)$$

Here, we predict that the tone of an analyst report has greater predictive power for both contemporaneous forecast errors and subsequent forecast revisions when the analyst is more constrained—a negative coefficient on $Tone \times Busyness$ in equation (11) and a positive coefficient on $Tone \times Busyness$ in equation (12).

Table 7 reports the results. In the $FError$ regressions, the estimated coefficient on $Tone \times Busyness$ is negative and significant at the 10% level with the large-portfolio proxy (column 1) and the 1% level with the busy-day proxy (column 2). In the $FRev_{t+1}$ regressions, it is positive and significant at the 10% level with the large-portfolio proxy (column 3) but insignificant with the busy-day proxy (column 4). Note that the busy-day specifications are estimated on prompt reports, for which concurrent disclosures constrain attention by construction (see footnote 21). The delayed-quantification channel should be strongest when the analyst has slack to catch up. Hence, in an untabulated analysis, we re-estimate the $FRev_{t+1}$ regression on the subsample in which the subsequent earnings forecast comes from a non-prompt report—one not issued concurrently with the firm's earnings announcements or management forecasts. Consistent with H2b, the estimated coefficient on $Tone \times Busyness$ is positive and significant at the 10% level (t-value = 1.94).

[21] This analysis uses a sample of prompt reports (defined as reports issued in the [-1, +1] window of the firm's earnings announcement or management forecast). These reports reflect situations where analysts face time pressure to respond to the new corporate announcements and promptly update their numerical forecasts.

*[Insert Table 7 here]*

Overall, the results support H1b and H2b that processing-capacity constraints reduce the extent to which analysts integrate their qualitative expectations into numerical forecasts, exacerbating quantification inefficiency.

## 6. Delineation from Alternative Explanations

We now empirically distinguish the quantification friction ($\theta$) from the strategic and behavioral alternatives outlined in Section 3.4; Table 8 reports three sets of tests.

First, according to our model, the quantification friction operates symmetrically on both positive and negative qualitative expectations. We therefore decompose $Tone$ in equations (7) and (9) into the percentage of positive sentences ($Pos\%$) and the percentage of negative sentences ($Neg\%$). Table 8, Panel A shows that both components are significant at least at the 5% level with the predicted signs; that is, positive text predicts negative forecast errors and upward revisions, and negative text predicts positive forecast errors and downward revisions. Because analysts underquantify good news as well as bad news, this symmetric underreaction supports the quantification-cost mechanism rather than intentional optimistic bias, under which predictability would be concentrated in negative signals.[22]

*[Insert Table 8 here]*

[22] Berger et al. (2019) find that analysts avoid quantifying good news (i.e., pessimistic bias) to provide management with beatable benchmarks near earnings announcements, whereas the quantification inefficiency we document appears at forecast horizons beyond one quarter (Table 6, Panel A).

Second, the structural quantification friction in our model should be uncorrelated with institutional catering incentives—delaying numerical updates to tip private clients with text—or reputational concerns—holding forecasts sticky to avoid seesaw revisions. We therefore re-estimate the main models in subsamples where such strategic incentives are weaker: (1) firms with low institutional ownership, (2) non-all-star analysts, and (3) analysts at independent research firms. Table 8, Panel B shows that the estimated coefficient on $Tone$ remains statistically and economically significant in all three subsamples, indicating that the observed inefficiency is not merely an artifact of strategic tipping to institutional clients or reputation management.

Finally, we address behavioral explanations such as conservatism or anchoring. Naive psychological errors attenuate as professional forecasters learn from experience, whereas a structural quantification friction should persist across experience levels. We partition the sample by the analyst's general experience and firm-specific experience. Table 8, Panel C shows that quantification inefficiency remains significant among analysts with extensive general and firm-specific experience. That is, seasoned analysts fail to eliminate the inefficiency through learning. These results indicate that the underreaction reflects a structural friction inherent in the integration of qualitative information rather than a transitory behavioral error.

## 7. Conclusion

A prevailing view in finance holds that any investment thesis "can almost always be distilled into outcomes that are objective, within a specific time horizon, and occur with an estimated probability."[23] Our study shows that this is neither always, nor even almost true. Many value-relevant signals in analyst reports have vague outcomes, unspecific horizons, or undefined probabilities, and quantifying them is costly. We argue that analysts leave such signals unquantified not because of strategic incentives or behavioral bias, but because integrating qualitative information into precise numbers is costly. We refer to the resulting outcome as quantification inefficiency: a systematic loss of value-relevant information that arises when quantification is costly.

[23] Michael Mauboussin and Dan Callahan, *Probabilities and payoffs: The practicalities and psychology of expected value*, Morgan Stanley, February 19, 2025, p.11.

To isolate integration costs from the costs of information awareness and acquisition, we examine the relation between the textual and numerical research outputs of the same analyst report. We find that textual tone negatively predicts contemporaneous forecast errors, indicating that qualitative insights are not fully captured in the quantitative output, and that tone positively predicts subsequent forecast revisions, indicating that analysts postpone quantification until further evidence resolves the underlying uncertainty. Furthermore, the inefficiency is exacerbated by linguistic vagueness, environmental uncertainty, and analyst busyness, confirming that quantification costs drive the observed inefficiency. These patterns persist where strategic incentives and behavioral biases are weaker, underscoring costly quantification as a structural feature of human information processing.

Collectively, our findings speak to the emerging “man plus machine” division of labor. Human analysts retain a durable advantage in acquiring qualitative signals in the physical world and articulating them in narrative form; machines, in turn, can harden these narratives into numbers at scale, relieving precisely the integration friction we document. We believe that by pinpointing which human information-processing friction technology can relieve, our study clarifies where the value of human judgment will persist.

# Appendix: Variable Definitions

## Textual Characteristics of Analyst Reports

$Tone$ — The textual tone of a report, calculated as the percentage of positive sentences minus the percentage of negative sentences, classified using FinBERT (Huang et al., 2023).

$TextOnly\%$ — The percentage of report sentences that contain neither "$" nor "%" (Huang et al., 2014).

$Hedge\%$ — The percentage of report sentences that contain hedging words. The wordlist is based on hedges in Lakoff (1973), approximators, vague quantifiers, and vague qualifiers in Channell (1994), and weak modal words and uncertain words in Loughran and McDonald (2011), tailored for the domain of analyst reports. Specifically, the wordlist includes the following terms (in lemmatized form):

Subjective beliefs:
- think, believe, feel, sense, suppose, suggest, argue
- in my/our view, from my/our perspective, as far as I/we can tell, to the best of my/our knowledge

Vague probability:
- seem, appear, apparent, sound, look like
- may, might, could, would, should
- maybe, perhaps, unlikely, improbable/improbably, potentially, possible/possibly, likely, probable/probably, conceivable/conceivably, presumably

Vague quantity, time, and frequency:
- around, approximate/approximately, roughly,
- few, bit, little, less, minority, some, several, number of, couple of, numerous, portion of, lot, mass, many, plenty, much, more, majority, most of
- sometime, earlier, recent/recently, soon, later
- seldom, sometimes, occasionally, often

Vague extent:
- sort of, kind of, more or less, slight/slightly
- fairly, pretty, relatively
- mostly, largely, principally, mainly, predominantly
- almost, nearly, practically, virtually, nominally, not entirely, close to
- as it were, so to say/speak, in a manner of speaking
- somewhat, to some degree/extent, to a certain degree/extent, to a large degree/extent

Vague manner:
- typical/typically, usually, in essence, essentially, in general, generally, basically, as a rule, tend to, apt to, prone to
- something, someone, somebody, somewhere, someplace, somehow, someway
- in some/most cases, in a/one sense, in a/one way

## Numerical Characteristics of Analyst Reports

| | |
|---|---|
| $FError$ | Earnings forecast error, calculated as an earnings forecast minus actual earnings from I/B/E/S, scaled by the stock price 50 days prior to the forecast date, and winsorized at the 1st and 99th percentiles. |
| $FRev_{t+1}$ | Subsequent earnings forecast revision, calculated as the subsequent earnings forecast minus the contemporaneous earnings forecast from I/B/E/S issued by the same analyst for the same firm and fiscal year, scaled by the stock price 50 days prior to the subsequent forecast date, and winsorized at the 1st and 99th percentiles. |
| $TPRev_{t+1}$ | Subsequent target price revision, calculated as the subsequent target price minus the contemporaneous target price from I/B/E/S issued by the same analyst for the same firm, scaled by the stock price 50 days prior to the subsequent target price announcement date, and winsorized at the 1st and 99th percentiles. |
| $RRev_{t+1}$ | Subsequent recommendation revisions, calculated as the subsequent stock recommendation (where high value indicates positive) minus the contemporaneous stock recommendation from I/B/E/S issued by the same analyst for the same firm. |
| $Horizon_{t\ (or\ t+1)}$ | The number of days between the forecast date (or the subsequent forecast date) and the annual earnings announcement date, winsorized at the 1st and 99th percentiles. |
| $Bold_{t\ (or\ t+1)}$ | Indicator variable equal to one if the forecast (or the subsequent forecast) deviates by more than two standard deviations from the consensus of forecasts in the prior 30 days, and zero otherwise. |
| $PriorCAR_{t\ (or\ t+1)}$ | The value-weighted market-adjusted two-day cumulative abnormal returns prior to the forecast (or the subsequent forecast, target price, or stock recommendation) date. |

## Analyst and Broker Characteristics

| | |
|---|---|
| $AnaGenExp$ | The number of years an analyst appears in I/B/E/S in a given year. |
| $AnaFirmExp$ | The number of years an analyst covers a firm in a given year. |
| $AnaAllStar$ | An indicator variable equal to one if an analyst is ranked as an All-Star by *Institutional Investor* in a given year, and zero otherwise. |
| $AnaLFR$ | The Leader-Follower Ratio (Cooper et al., 2001) of an analyst in a given year. |
| $AnaNFirm$ | The number of firms that an analyst covers in a given year. |
| $AnaNFore$ | The number of total forecasts that an analyst issues in a given year. |
| $BHSize$ | The number of analysts in a brokerage house in a given year. |
| $BHIB$ | An indicator variable equal to one if a brokerage house is an investment bank, and zero otherwise. |

## Firm Characteristics

| | |
|---|---|
| $FirmSize$ | The log value of a firm's market value of equity at year end. |
| $FirmBM$ | The book-to-market ratio of equity of a firm at year end. |
| $FirmIO$ | The percentage of shares outstanding owned by institutions at the year end. |
| $FirmNAna$ | The number of analysts covering a firm in a given year. |
| $FirmNFore$ | The number of forecasts for a firm in a given year. |

## Figure 1. Tone and Forecast Errors over Time

This figure plots mean earnings forecast errors ($FError$, measured as the earnings forecast minus actual earnings) across quintiles of analyst report $Tone$. Q1 is the most pessimistic quintile and Q5 is the most optimistic quintile. A positive (negative) forecast error means that the forecast is optimistically (pessimistically) biased. The figure displays the mean forecast errors for the current report $t$ and up to four subsequent reports from $t + 1$ to $t + 4$ for the same fiscal year.

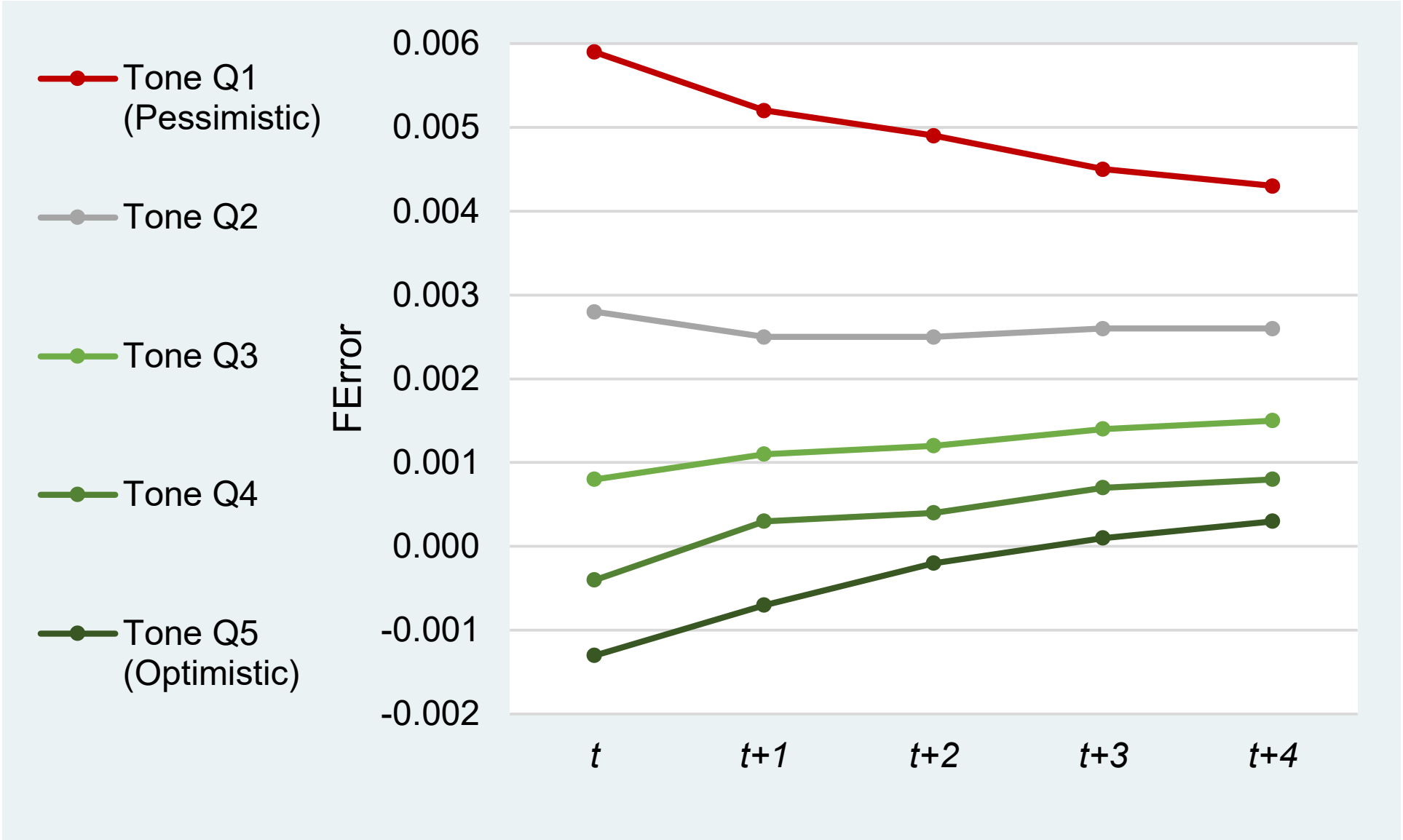

## Table 1. Sample Selection

This table describes the selection of analyst reports from Investext and the matching process with I/B/E/S data.

| | |
|---|---|
| Reports for S&P 500 firms from 1995 to 2012, obtained from Investext | 635,041 |
| Less: Reports without matching analyst or firm identifiers in I/B/E/S | (162,772) |
| Reports matched with I/B/E/S identifiers | 472,269 |
| Less: Reports without any contemporaneous numerical forecasts | (231,224) |
| Final sample: Reports with at least one numerical forecast | 241,045 |
| Reports with earnings forecasts | 209,723 |
| Reports with target prices | 88,849 |
| Reports with stock recommendations | 24,986 |

## Table 2. Descriptive Statistics

This table presents the descriptive statistics for the main variables used in this study. The Appendix provides detailed definitions of the variables. N is the number of observations, and Mean and SD are sample means and standard deviations of variables, respectively. Min, 25th, 50th, 75th, and Max show the sample distribution of the variables.

| Variable | N | Mean | SD | Min | 25th | 50th | 75th | Max |
|---|---|---|---|---|---|---|---|---|
| $Tone$ | 241,045 | 0.14 | 0.31 | -1.00 | -0.06 | 0.16 | 0.35 | 1.00 |
| $TextOnly\%$ | 241,045 | 0.57 | 0.17 | 0.00 | 0.45 | 0.57 | 0.68 | 1.00 |
| $Hedge\%$ | 241,045 | 0.38 | 0.15 | 0.00 | 0.29 | 0.38 | 0.48 | 1.00 |
| | | | | | | | | |
| $FError$ | 209,723 | 0.0017 | 0.0188 | -0.0532 | -0.0030 | -0.0004 | 0.0024 | 0.1160 |
| $FRev_{t+1}$ | 157,510 | -0.0007 | 0.0081 | -0.0445 | -0.0018 | 0.0003 | 0.0017 | 0.0262 |
| $Bold$ | 180,793 | 0.31 | 0.46 | 0.00 | 0.00 | 0.00 | 1.00 | 1.00 |
| $Horizon$ | 209,723 | 207.40 | 103.84 | 8.00 | 105.00 | 195.00 | 287.00 | 371.00 |
| $PriorCAR$ | 209,716 | 0.00 | 0.04 | -0.94 | -0.02 | 0.00 | 0.02 | 1.06 |
| | | | | | | | | |
| $TPRev_{t+1}$ | 88,849 | 0.00 | 0.21 | -0.85 | -0.08 | 0.03 | 0.11 | 0.55 |
| $RRev_{t+1}$ | 24,986 | 0.00 | 1.25 | -4.00 | -1.00 | 0.00 | 1.00 | 4.00 |
| | | | | | | | | |
| $AnaGenExp$ | 209,733 | 9.11 | 5.83 | 1.00 | 5.00 | 8.00 | 12.00 | 31.00 |
| $AnaFirmExp$ | 209,733 | 5.49 | 4.54 | 1.00 | 2.00 | 4.00 | 7.00 | 31.00 |
| $AnaAllStar$ | 209,733 | 0.22 | 0.42 | 0.00 | 0.00 | 0.00 | 0.00 | 1.00 |
| $AnaLFR$ | 185,034 | 2.00 | 3.01 | 0.00 | 0.32 | 1.00 | 2.28 | 18.25 |
| $AnaNFirm$ | 209,733 | 15.42 | 6.50 | 1.00 | 11.00 | 15.00 | 19.00 | 109.00 |
| $AnaNFore$ | 209,733 | 73.56 | 43.00 | 1.00 | 45.00 | 67.00 | 92.00 | 454.00 |
| $BHSize$ | 209,733 | 79.67 | 51.17 | 1.00 | 33.00 | 66.00 | 120.00 | 217.00 |
| $BHIB$ | 209,733 | 0.73 | 0.44 | 0.00 | 0.00 | 1.00 | 1.00 | 1.00 |
| | | | | | | | | |
| $FirmSize$ | 208,411 | 9.55 | 1.15 | 6.94 | 8.78 | 9.46 | 10.26 | 12.43 |
| $FirmBM$ | 178,672 | 0.51 | 0.37 | 0.05 | 0.25 | 0.41 | 0.66 | 2.04 |
| $FirmIO$ | 188,194 | 0.74 | 0.17 | 0.00 | 0.65 | 0.76 | 0.85 | 1.00 |
| $FirmNAna$ | 209,733 | 26.73 | 10.06 | 1.00 | 20.00 | 25.00 | 33.00 | 70.00 |
| $FirmNFore$ | 209,733 | 128.95 | 70.16 | 1.00 | 78.00 | 114.00 | 167.00 | 494.00 |

## Table 3. Tone and Forecast Errors

This table presents the results of regressing an analyst's earnings forecast error ($FError$, measured as the earnings forecast minus actual earnings) on the $Tone$ of the analyst's contemporaneous report text and control variables. The Appendix provides detailed variable definitions. Fixed effects are untabulated, and observations with missing control variables or singleton fixed effects are dropped prior to estimation. Standard errors are two-way clustered by analyst and year. *t-statistics* are reported in parentheses. *, **, and *** indicate significance at the 10%, 5%, and 1% levels, respectively.

| | (1)<br>$FError$ | (2)<br>$FError$ | (3)<br>$FError$ |
|---|---|---|---|
| $Tone$ | -0.0038*** | -0.0013*** | -0.0011*** |
| | (-6.77) | (-3.75) | (-3.11) |
| $Horizon$ | 0.0000** | 0.0000* | 0.0000* |
| | (2.14) | (1.86) | (1.65) |
| $Bold$ | 0.0003 | -0.0005** | -0.0006** |
| | (0.90) | (-2.03) | (-2.33) |
| $PriorCAR$ | -0.0134*** | -0.0103*** | -0.0070** |
| | (-2.58) | (-3.35) | (-2.18) |
| $AnaGenExp$ | 0.0000 | | 0.0082* |
| | (1.18) | | (1.89) |
| $AnaFirmExp$ | 0.0000 | | -0.0027*** |
| | (0.87) | | (-13.01) |
| $AnaAllStar$ | -0.0001 | | 0.0005 |
| | (-0.31) | | (0.66) |
| $AnaLFR$ | -0.0000 | | 0.0000 |
| | (-1.04) | | (0.26) |
| $AnaNFirm$ | 0.0000 | | -0.0001 |
| | (1.18) | | (-0.84) |
| $AnaNFore$ | -0.0000* | | 0.0000 |
| | (-1.76) | | (0.38) |
| $BHSize$ | -0.0000 | | 0.0000 |
| | (-0.61) | | (0.11) |
| $BHIB$ | 0.0002 | | 0.0012 |
| | (0.60) | | (1.05) |
| $FirmSize$ | -0.0009*** | -0.0006** | -0.0042*** |
| | (-3.40) | (-2.24) | (-3.88) |
| $FirmBM$ | 0.0089*** | 0.0081*** | 0.0150*** |
| | (3.79) | (5.48) | (8.37) |
| $FirmIO$ | -0.0012 | 0.0001 | -0.0015 |
| | (-0.60) | (0.07) | (-0.67) |
| $FirmNAna$ | -0.0000 | -0.0001* | -0.0002** |
| | (-0.67) | (-1.78) | (-2.08) |
| $FirmNFore$ | 0.0000 | 0.0000** | 0.0001*** |
| | (0.15) | (2.55) | (3.47) |
| Constant | 0.0053 | 0.0012 | -0.0288 |
| | (1.31) | (0.27) | (-0.69) |
| Fixed Effects | Year | Analyst-year | Analyst-firm, Year |
| Adj. $R^2$ | 0.097 | 0.325 | 0.370 |
| # of Obs. | 122,268 | 135,677 | 120,103 |

## Table 4. Vaguer Reports

This table presents the variation in incomplete quantification based on the degree of linguistic vagueness across reports. An indicator variable for vaguer reports ($Vagueness$) and its interaction with tone ($Tone \times Vagueness$) are added to the regression models. In column 1, $Vagueness$ indicates that the percentage of report sentences containing neither "$" nor "%" ($TextOnly\%$) exceeds its median across the analyst's own reports. In column 2, $Vagueness$ indicates that the percentage of report sentences containing hedging words ($Hedge\%$) exceeds its median across the analyst's own reports. For the sake of brevity, only the coefficients of interest are presented. Control variables are the same as in Table 3. Fixed effects are untabulated, and observations with missing control variables or singleton fixed effects are dropped prior to estimation. Standard errors are two-way clustered by analyst and year. *t-statistics* are reported in parentheses. *, **, and *** indicate significance at the 10%, 5%, and 1% levels, respectively.

| | Absence of Numbers<br>(1)<br>$FError$ | Use of Hedging Words<br>(2)<br>$FError$ |
|---|---|---|
| $Tone$ | -0.0008** | -0.0005 |
| | (-2.26) | (-1.06) |
| $Tone \times Vagueness$ | -0.0008** | -0.0014*** |
| | (-2.57) | (-4.07) |
| $Vagueness$ | 0.0001 | 0.0004** |
| | (0.49) | (2.55) |
| Control Variables | Yes | Yes |
| Analyst-firm FEs | Yes | Yes |
| Year FEs | Yes | Yes |
| Adj. $R^2$ | 0.370 | 0.370 |
| # of Obs. | 120,102 | 120,102 |

## Table 5. Tone and Subsequent Revisions

This table presents the results of regressing the analyst's subsequent revisions in numerical forecasts on the $Tone$ of the analyst's contemporaneous report text and control variables. Panel A reports the results of earnings forecast revisions ($FRev_{t+1}$, measured as the next forecast of the analyst for the same firm and fiscal year minus the contemporaneous forecast). Panel B reports the results of target price revisions ($TPRev_{t+1}$), excluding reports from Morgan Stanley, which provides range forecasts of target prices. Panel C reports the results of stock recommendation revisions ($RRev_{t+1}$), estimated separately based on the current recommendation level (i.e., Buy/Strong Buy, Hold, and Sell/Strong Sell, respectively). The Appendix provides detailed variable definitions. For the sake of brevity, only the coefficients of interest are presented in Panels B and C. The control variables in these two panels are the same as in Panel A (excluding $Horizon$ and $Bold$). Fixed effects are untabulated, and observations with missing control variables or singleton fixed effects are dropped prior to estimation. Standard errors are two-way clustered by analyst and year. *t-statistics* are reported in parentheses. *, **, *** indicate the significance at the 10%, 5%, and 1% levels, respectively.

### Panel A. Earnings Forecast Revisions

| | (1)<br>$FRev_{t+1}$ | (2)<br>$FRev_{t+1}$ | (3)<br>$FRev_{t+1}$ |
|---|---|---|---|
| $Tone$ | 0.0016*** | 0.0012*** | 0.0014*** |
| | (6.61) | (5.01) | (5.17) |
| $FRev_t$ | 0.1269*** | 0.0303** | 0.0095 |
| | (6.71) | (2.04) | (0.64) |
| $Horizon_{t+1}$ | 0.0000 | 0.0000 | 0.0000 |
| | (1.08) | (1.07) | (1.32) |
| $Bold_{t+1}$ | -0.0016*** | -0.0014*** | -0.0015*** |
| | (-4.87) | (-4.14) | (-4.31) |
| $PriorCAR_{t+1}$ | 0.0206*** | 0.0197*** | 0.0193*** |
| | (10.26) | (10.27) | (10.72) |
| $AnaGenExp$ | -0.0000 | | 0.0003 |
| | (-0.44) | | (0.99) |
| $AnaFirmExp$ | -0.0000** | | 0.0008*** |
| | (-2.41) | | (14.39) |
| $AnaAllStar$ | -0.0000 | | -0.0002 |
| | (-0.33) | | (-1.19) |
| $AnaLFR$ | 0.0000 | | 0.0000 |
| | (0.46) | | (0.71) |
| $AnaNFirm$ | -0.0000* | | -0.0000* |
| | (-1.81) | | (-1.68) |
| $AnaNFore$ | 0.0000** | | 0.0000 |
| | (2.32) | | (0.13) |
| $BHSize$ | 0.0000 | | -0.0000 |
| | (0.57) | | (-0.12) |
| $BHIB$ | 0.0000 | | -0.0004 |
| | (0.55) | | (-1.46) |
| $FirmSize$ | 0.0004*** | 0.0004*** | 0.0015*** |
| | (5.71) | (3.99) | (5.20) |
| $FirmBM$ | -0.0022*** | -0.0023*** | -0.0039*** |
| | (-4.58) | (-5.77) | (-4.80) |
| $FirmIO$ | 0.0011*** | 0.0009* | 0.0017 |
| | (2.67) | (1.83) | (1.39) |
| $FirmNAna$ | -0.0000 | 0.0000 | 0.0000* |
| | (-0.49) | (1.55) | (1.65) |
| $FirmNFore$ | 0.0000 | -0.0000** | -0.0000*** |
| | (0.40) | (-2.44) | (-4.29) |
| Constant | -0.0044*** | -0.0037*** | -0.0208*** |
| | (-5.11) | (-3.46) | (-6.17) |
| Fixed Effects | Year | Analyst-year | Analyst-firm, Year |
| Adj. $R^2$ | 0.089 | 0.158 | 0.164 |
| # of Obs. | 111,178 | 120,855 | 109,020 |

Panel B. Target Price Revisions

| | (1)<br>$TPRev_{t+1}$ | (2)<br>$TPRev_{t+1}$ | (3)<br>$TPRev_{t+1}$ |
|---|---|---|---|
| $Tone$ | 0.0315*** | 0.0254*** | 0.0262*** |
| | (4.51) | (3.84) | (4.27) |
| Control Variables | Yes | Yes | Yes |
| Fixed Effects | Year | Analyst-year | Analyst-firm, Year |
| Adj. $R^2$ | 0.203 | 0.275 | 0.251 |
| # of Obs. | 38,676 | 42,603 | 37,085 |

Panel C. Stock Recommendation Revisions

| Current Buy Reports | (1)<br>$RRev_{t+1}$ | (2)<br>$RRev_{t+1}$ | (3)<br>$RRev_{t+1}$ |
|---|---|---|---|
| $Tone$ | 0.3037*** | 0.2419*** | 0.2753*** |
| | (5.96) | (4.73) | (3.66) |
| Control Variables | Yes | Yes | Yes |
| Fixed Effects | Year | Analyst-year | Analyst-firm, Year |
| Adj. $R^2$ | 0.140 | 0.553 | 0.428 |
| # of Obs. | 3,498 | 1,849 | 1,294 |

| Current Hold Reports | (4)<br>$RRev_{t+1}$ | (5)<br>$RRev_{t+1}$ | (6)<br>$RRev_{t+1}$ |
|---|---|---|---|
| $Tone$ | 0.5270*** | 0.2895*** | 0.3235*** |
| | (10.84) | (7.83) | (3.61) |
| Control Variables | Yes | Yes | Yes |
| Fixed Effects | Year | Analyst-year | Analyst-firm, Year |
| Adj. $R^2$ | 0.150 | 0.507 | 0.528 |
| # of Obs. | 3,095 | 1,530 | 1,082 |

| Current Sell Reports | (7)<br>$RRev_{t+1}$ | (8)<br>$RRev_{t+1}$ | (9)<br>$RRev_{t+1}$ |
|---|---|---|---|
| $Tone$ | -0.1253* | -0.0424 | 0.1200* |
| | (-1.74) | (-0.28) | (1.83) |
| Control Variables | Yes | Yes | Yes |
| Fixed Effects | Year | Analyst-year | Analyst-firm, Year |
| Adj. $R^2$ | 0.066 | 0.591 | 0.631 |
| # of Obs. | 678 | 209 | 129 |

## Table 6. Heightened Uncertainty

This table presents the variation in delayed quantification based on the degree of uncertainty over time and across firms. In Panel A, the results are estimated separately within the four groups of forecast horizons partitioned by quarter. In Panels B through E, an indicator variable for heightened uncertainty ($Uncertainty$) and its interaction with tone ($Tone \times Uncertainty$) are added to the regression models. In Panel B, $Uncertainty$ indicates that a report is announced (1) in months where the macroeconomic uncertainty index of Jurado et al. (2015) exceeds the sample median, (2) during recession periods as identified by the National Bureau of Economic Research, or (3) in months where the Chicago Board Options Exchange's volatility index (VIX) exceeds the sample median. In Panel C, $Uncertainty$ indicates that (1) a firm's five-year earnings per share volatility exceeds the sample median, or (2) a firm's one-year daily stock return volatility exceeds the sample median. In Panel D, $Uncertainty$ indicates that a firm (1) undergoes a merger or acquisition during the year, (2) changes its number of reported business segments during the year, or (3) experiences CEO or CFO turnover during the year. In Panel E, $Uncertainty$ indicates that (1) firm size is below the sample median, (2) analyst coverage is below the sample median, or (3) the analyst forecast dispersion exceeds the sample median. For the sake of brevity, only the coefficients of interest are presented. Control variables are the same as in Panel A of Table 5. Fixed effects are untabulated, and observations with missing control variables or singleton fixed effects are dropped prior to estimation. Standard errors are two-way clustered by analyst and year. *t-statistics* are reported in parentheses. *, **, and *** indicate significance at the 10%, 5%, and 1% levels, respectively.

### Panel A. Long Forecast Horizons

| Forecast Horizons | 0 to 89 days<br>(1)<br>$FRev_{t+1}$ | 90 to 179 days<br>(2)<br>$FRev_{t+1}$ | 180 to 269 days<br>(3)<br>$FRev_{t+1}$ | 270 to 360 days<br>(4)<br>$FRev_{t+1}$ |
|---|---|---|---|---|
| $Tone$ | -0.0002 | 0.0006*** | 0.0014*** | 0.0018*** |
| | (-0.23) | (4.03) | (4.08) | (8.24) |
| Control Variables | Yes | Yes | Yes | Yes |
| Analyst-firm FEs | Yes | Yes | Yes | Yes |
| Year FEs | Yes | Yes | Yes | Yes |
| Adj. $R^2$ | 0.365 | 0.254 | 0.257 | 0.193 |
| # of Obs. | 2,367 | 21,405 | 30,371 | 30,361 |

### Panel B. High Macroeconomic and Market Volatility

| | Macro Uncertainty<br>(1)<br>$FRev_{t+1}$ | Recession<br>(2)<br>$FRev_{t+1}$ | VIX<br>(3)<br>$FRev_{t+1}$ |
|---|---|---|---|
| $Tone$ | 0.0007** | 0.0010*** | 0.0011*** |
| | (2.22) | (4.87) | (5.78) |
| $Tone \times Uncertainty$ | 0.0013*** | 0.0015** | 0.0007** |
| | (2.76) | (2.31) | (1.97) |
| $Uncertainty$ | -0.0004* | -0.0027*** | -0.0005** |
| | (-1.67) | (-3.75) | (-2.41) |
| Control Variables | Yes | Yes | Yes |
| Analyst-firm FEs | Yes | Yes | Yes |
| Year FEs | Yes | Yes | Yes |
| Adj. $R^2$ | 0.164 | 0.167 | 0.164 |
| # of Obs. | 109,020 | 109,020 | 108,927 |

Panel C. High Firm-level Volatility

| | Earnings Volatility (1) $FRev_{t+1}$ | Return Volatility (2) $FRev_{t+1}$ |
|---|---|---|
| $Tone$ | 0.0010*** | 0.0005** |
| | (5.30) | (2.43) |
| $Tone \times Uncertainty$ | 0.0009** | 0.0018*** |
| | (2.23) | (5.52) |
| $Uncertainty$ | -0.0001 | -0.0003 |
| | (-0.65) | (-1.05) |
| Control Variables | Yes | Yes |
| Analyst-firm FEs | Yes | Yes |
| Year FEs | Yes | Yes |
| Adj. $R^2$ | 0.164 | 0.166 |
| # of Obs. | 108,411 | 107,031 |

Panel D. Significant Corporate Changes

| | Mergers and Acquisitions (1) $FRev_{t+1}$ | Changes in Segments (2) $FRev_{t+1}$ | CEO and CFO Turnovers (3) $FRev_{t+1}$ |
|---|---|---|---|
| $Tone$ | 0.0011*** | 0.0012*** | 0.0009*** |
| | (3.65) | (4.11) | (3.09) |
| $Tone \times Uncertainty$ | 0.0007*** | 0.0013** | 0.0008*** |
| | (2.93) | (2.23) | (2.90) |
| $Uncertainty$ | -0.0004** | -0.0005** | -0.0006*** |
| | (-2.44) | (-2.13) | (-4.89) |
| Control Variables | Yes | Yes | Yes |
| Analyst-firm FEs | Yes | Yes | Yes |
| Year FEs | Yes | Yes | Yes |
| Adj. $R^2$ | 0.164 | 0.155 | 0.164 |
| # of Obs. | 109,022 | 86,470 | 109,022 |

Panel E. Poor Information Environments

| | Small Firm Size (1) $FRev_{t+1}$ | Low Analyst Coverage (2) $FRev_{t+1}$ | High Forecast Dispersion (3) $FRev_{t+1}$ |
|---|---|---|---|
| $Tone$ | 0.0008** | 0.0010*** | 0.0002** |
| | (2.57) | (3.33) | (2.08) |
| $Tone \times Uncertainty$ | 0.0013*** | 0.0009*** | 0.0018*** |
| | (3.68) | (3.42) | (6.05) |
| $Uncertainty$ | 0.0004 | 0.0002 | -0.0002 |
| | (1.33) | (1.37) | (-0.92) |
| Control Variables | Yes | Yes | Yes |
| Analyst-firm FEs | Yes | Yes | Yes |
| Year FEs | Yes | Yes | Yes |
| Adj. $R^2$ | 0.164 | 0.164 | 0.172 |
| # of Obs. | 109,020 | 109,020 | 91,889 |

## Table 7. Busy Analysts

This table presents the variation in quantification inefficiency based on the degree of workload across analysts. An indicator variable for busy analysts ($Busyness$) and its interaction with tone ($Tone \times Busyness$) are added to the regression models. In columns 1 and 3, $Busyness$ indicates that the number of firms covered by an analyst exceeds the sample median. In columns 2 and 4, $Busyness$ indicates that multiple firms in the analyst's portfolio issue an earnings announcement or a management forecast in the [-1, +1] window of the report release date; this analysis restricts the sample to prompt reports, defined as those issued in the [-1, +1] windows of the firm's earnings announcements or management forecasts. For the sake of brevity, only the coefficients of interest are presented. Control variables are the same as in Table 3 and Panel A of Table 5. Fixed effects are untabulated, and observations with missing control variables or singleton fixed effects are dropped prior to estimation. Standard errors are two-way clustered by analyst and year. *t-statistics* are reported in parentheses. *, **, and *** indicate significance at the 10%, 5%, and 1% levels, respectively.

| | Large Portfolio (1) $FError$ | Busy Day (2) $FError$ |
|---|---|---|
| $Tone$ | -0.0005 | -0.0009** |
| | (-1.47) | (-2.40) |
| $Tone \times Busyness$ | -0.0010* | -0.0012*** |
| | (-1.83) | (-2.71) |
| $Busyness$ | 0.0004 | 0.0004* |
| | (0.90) | (1.77) |
| Control Variables | Yes | Yes |
| Analyst-firm FEs | Yes | Yes |
| Year FEs | Yes | Yes |
| Adj. $R^2$ | 0.370 | 0.369 |
| # of Obs. | 120,109 | 68,767 |

| | Large Portfolio (3) $FRev_{t+1}$ | Busy Day (4) $FRev_{t+1}$ |
|---|---|---|
| $Tone$ | 0.0012*** | 0.0014*** |
| | (4.47) | (4.46) |
| $Tone \times Busyness$ | 0.0003* | 0.0002 |
| | (1.71) | (0.87) |
| $Busyness$ | -0.0001 | -0.0002 |
| | (-0.54) | (-1.45) |
| Control Variables | Yes | Yes |
| Analyst-firm FEs | Yes | Yes |
| Year FEs | Yes | Yes |
| Adj. $R^2$ | 0.164 | 0.191 |
| # of Obs. | 109,017 | 66,036 |

## Table 8. Tests of Alternative Explanations

This table presents results addressing alternative explanations for the main findings. In Panel A, the percentages of positive sentences ($Pos\%$) and negative sentences ($Neg\%$) replace *Tone* in the regression models. In Panel B, results are estimated using subsamples of firms with below-median institutional ownership within an analyst's portfolio (columns 1 and 4), analysts not ranked as All-Stars (columns 2 and 5), and analysts at independent research firms (columns 3 and 6). In Panel C, results are estimated using subsamples of analysts whose general experience (columns 1 and 3) or firm-specific experience (columns 2 and 4) exceeds the annual median. For the sake of brevity, only the coefficients of interest are presented. Control variables for $FError$ and $FRev_{t+1}$ are the same as in Table 3 and Panel A of Table 5, respectively. Fixed effects are untabulated, and observations with missing control variables or singleton fixed effects are dropped prior to estimation. Standard errors are two-way clustered by analyst and year. *t-statistics* are reported in parentheses. *, **, and *** indicate significance at the 10%, 5%, and 1% levels, respectively.

### Panel A. Optimistic and Pessimistic Bias

| | (1)<br>$FError$ | (2)<br>$FRev_{t+1}$ |
|---|---|---|
| $Pos\%$ | -0.0012** | 0.0010*** |
| | (-2.20) | (2.66) |
| $Neg\%$ | 0.0010** | -0.0020*** |
| | (2.31) | (-8.03) |
| Control Variables | Yes | Yes |
| Analyst-firm FEs | Yes | Yes |
| Year FEs | Yes | Yes |
| Adj. $R^2$ | 0.370 | 0.164 |
| # of Obs. | 120,109 | 109,017 |

### Panel B. Weaker Institutional Catering Incentives and Reputational Concerns

| | Low Institutional Ownership<br>(1)<br>$FError$ | Non-All-Star Analysts<br>(2)<br>$FError$ | Independent Research Firms<br>(3)<br>$FError$ |
|---|---|---|---|
| $Tone$ | -0.0007* | -0.0012*** | -0.0053** |
| | (-1.73) | (-3.15) | (-2.02) |
| Control Variables | Yes | Yes | Yes |
| Analyst-year FEs | Yes | Yes | No |
| Year FEs | No | No | Yes |
| Adj. $R^2$ | 0.369 | 0.325 | 0.066 |
| # of Obs. | 66,429 | 104,902 | 1,185 |

| | Low Institutional Ownership<br>(4)<br>$FRev_{t+1}$ | Non-All-Star Analysts<br>(5)<br>$FRev_{t+1}$ | Independent Research Firms<br>(6)<br>$FRev_{t+1}$ |
|---|---|---|---|
| $Tone$ | 0.0012*** | 0.0012*** | 0.0041*** |
| | (4.13) | (4.50) | (3.57) |
| Control Variables | Yes | Yes | Yes |
| Analyst-year FEs | Yes | Yes | No |
| Year FEs | No | No | Yes |
| Adj. $R^2$ | 0.191 | 0.153 | 0.032 |
| # of Obs. | 59,727 | 92,514 | 882 |

Panel C. Weaker Behavioral Biases

| | High General Experience (1) $FError$ | High Firm Experience (2) $FError$ | High General Experience (3) $FRev_{t+1}$ | High Firm Experience (4) $FRev_{t+1}$ |
|---|---|---|---|---|
| $Tone$ | -0.0016*** | -0.0013*** | 0.0011*** | 0.0010*** |
| | (-3.86) | (-3.29) | (5.02) | (4.45) |
| Control Variables | Yes | Yes | Yes | Yes |
| Analyst-year FEs | Yes | Yes | Yes | Yes |
| Adj. $R^2$ | 0.314 | 0.374 | 0.167 | 0.176 |
| # of Obs. | 64,081 | 61,903 | 57,343 | 56,214 |

## Internet Appendix

## Table IA1. Robustness Tests

This table presents the results of the robustness tests described in footnotes 17 and 18. Panel A replaces the two-day prior returns ($PriorCAR$) with a two-month prior returns ($2MPriorCAR$), adds prior annual earnings surprises ($ESurp$), and controls for contemporaneous revisions ($FRev_t$) in the $FError$ regression. Panel B partitions the sample into prompt and non-prompt reports, defined as reports issued in or out the [-1, +1] window of the firm's earnings announcement or management forecast, respectively. Panel C employs the alternative tone measures: the change in $Tone$ relative to the analyst's prior report and an analyst-firm-demeaned $Tone$. Panel D tests the predictability power of an analyst's current report $Tone$ on the earnings forecast revisions across multiple subsequent reports. For the sake of brevity, only the coefficients of interest are presented. Except in Panel A, control variables for $FError$ and $FRev_{t+1\ (t+2,t+3,t+4)}$ are the same as in Table 3 and Panel A of Table 5, respectively. Fixed effects are untabulated, and observations with missing control variables or singleton fixed effects are dropped prior to estimation. Standard errors are two-way clustered by analyst and year. *t-statistics* are reported in parentheses. *, **, and *** indicate significance at the 10%, 5%, and 1% levels, respectively.

Panel A. Alternative Control Variables

| | (1)<br>$FError$ | (2)<br>$FRev_{t+1}$ |
|---|---|---|
| $Tone$ | -0.0007** | 0.0010*** |
| | (-2.18) | (3.03) |
| $FRev_t$ | -0.0372 | -0.0015 |
| | (-0.77) | (-0.10) |
| $2MPriorCAR_{t\ or\ t+1}$ | -0.0035** | 0.0075*** |
| | (-2.12) | (7.53) |
| $ESurp$ | -0.0229** | 0.0040 |
| | (-2.08) | (1.10) |
| Control Variables | Yes | Yes |
| Analyst-firm FEs | Yes | Yes |
| Year FEs | Yes | Yes |
| Adj. $R^2$ | 0.370 | 0.173 |
| # of Obs. | 105,388 | 108,821 |

Panel B. Prompt versus Non-prompt Reports

| | Prompt Reports | | Non-prompt Reports | |
|---|---|---|---|---|
| | (1)<br>$FError$ | (2)<br>$FRev_{t+1}$ | (3)<br>$FError$ | (4)<br>$FRev_{t+1}$ |
| $Tone$ | -0.0012*** | 0.0015*** | -0.0014*** | 0.0015*** |
| | (-3.24) | (4.50) | (-3.07) | (5.31) |
| Control Variables | Yes | Yes | Yes | Yes |
| Analyst-firm FEs | Yes | Yes | Yes | Yes |
| Year FEs | Yes | Yes | Yes | Yes |
| Adj. $R^2$ | 0.369 | 0.191 | 0.354 | 0.154 |
| # of Obs. | 68,767 | 66,036 | 47,531 | 39,334 |

Panel C. Alternative Tone Measures

| | Tone Revisions | | Demeaned Tone | |
|---|---|---|---|---|
| | (1) | (2) | (3) | (4) |
| | $FError$ | $FRev_{t+1}$ | $FError$ | $FRev_{t+1}$ |
| $Tone$ | -0.0006*** | 0.0003*** | -0.0011*** | 0.0014*** |
| | (-3.89) | (2.68) | (-3.12) | (5.17) |
| Control Variables | Yes | Yes | Yes | Yes |
| Analyst-firm FEs | Yes | Yes | Yes | Yes |
| Year FEs | Yes | Yes | Yes | Yes |
| Adj. $R^2$ | 0.373 | 0.164 | 0.370 | 0.163 |
| # of Obs. | 112,781 | 102,918 | 120,109 | 109,017 |

Panel D. Revisions Across Multiple Subsequent Reports

| | (1) | (2) | (3) |
|---|---|---|---|
| | $FRev_{t+2}$ | $FRev_{t+3}$ | $FRev_{t+4}$ |
| $Tone$ | 0.0007*** | 0.0005*** | 0.0000 |
| | (5.00) | (3.37) | (0.06) |
| Control Variables | Yes | Yes | Yes |
| Analyst-firm FEs | Yes | Yes | Yes |
| Year FEs | Yes | Yes | Yes |
| Adj. $R^2$ | 0.163 | 0.180 | 0.196 |
| # of Obs. | 82,420 | 58,936 | 39,836 |